\documentstyle[aps,prb,twocolumn,epsfig]{revtex}  
\begin{document}
\twocolumn[\hsize\textwidth\columnwidth\hsize\csname @twocolumnfalse\endcsname
\title{Bond-order and charge-density waves in the isotropic interacting
  two-dimensional quarter-filled band and the insulating state proximate to
  organic superconductivity}

\author{S.~Mazumdar$^1$, R.~T.~Clay$^{1,2}$, and D.~K.~Campbell$^3$}
\address{$^1$ Department of Physics, University of Arizona,
Tucson, AZ 85721 \\ $^2$ Cooperative Excitation Project ERATO, Japan Science and
Technology Corporation (JST) \\$^3$ Department of Physics, University of Illinois
,
Urbana, Illinois 61801}

\date{\today}

\maketitle
\begin{abstract}

We report three surprising results regarding the nature of the spatial broken
symmetries in the two-dimensional (2D),
quarter-filled band with strong electron-electron interactions that provides a
microscopic model of the
2:1 cationic organic charge transfer solids (CTS). First,
in direct contradiction to the predictions of one-electron theory, we find
a coexisting  ``bond-order and charge-density wave'' (BCDW) insulating ground
state in the 2D rectangular lattice for {\it all} anisotropies,
including the isotropic limit. Second, in contrast to the interacting
half-filled band, which exhibits one singlet-to-antiferromagnet (AFM) 
transition as the
interchain coupling is increased from zero, there occur in the interacting
quarter-filled band two distinct transitions: a similar singlet-to-
antiferromagnet/spin-density
wave (AFM/SDW) transition at small interchain coupling, giving rise to a
bond-charge-spin density wave (BCSDW) state, followed by a second
AFM/SDW-to-singlet transition at large interchain coupling. Third, we show
that our conclusions remain unchanged if one assumes
the conventional ``effective 1/2-filled''
lattice of dimer sites for the CTS: the dimer lattice unconditionally
dimerizes again to give the same BCDW found in the quarter-filled band.
We make detailed comparisons to recent experiments
in the tetramethyl-tetrathiafulvalene (TMTTF),        tetramethyl-tetraselenafulvalene (TMTSF),
bisethylenedithio-tetrathiafulvalene (BEDT-TTF) and
bisethylenedithio-tetraselenafulvalene (BETS)-based CTS.
Our theory
explains the mixed
charge-spin density waves observed in TMTSF and certain BEDT-TTF systems,
as well
as the
absence of antiferromagnetism in the BETS-based systems.
An important consequence of this work is the suggestion
that organic superconductivity is related to the
proximate Coulomb-induced BCDW, with the SDW that coexists for
large anisotropies being also a consequence of the BCDW, rather than the
driver of superconductivity. We point out that the BCDW and BCSDW states are
analogous to four different classes of ``paired'' semiconductors that
are obtained within certain models of exotic superconductivity.
That all four of these models can in principle give rise to superconductivity
in the weakly incommensurate regime provides further motivation for
the notion that the BCDW may be driving
the superconductivity in the organics.    
\end{abstract}

]

\section{Introduction}

Theoretical discussions of spatial broken symmetries in strongly correlated
electron systems have largely focused on the 1/2-filled band Mott-Hubbard 
semiconductor. The one-dimensional (1D) case has been widely discussed in the
context of polyacetylene \cite{review1,review2}. Here it is known that Coulomb 
electron-electron (e-e) interactions can strongly enhance the  
2k$_F$ 
(k$_F$ = one-electron
Fermi wavevector) bond-alternation expected within the Peierls
purely electron-phonon (e-ph) coupled model, giving
rise to a periodic modulation of the bond-order, a bond-order wave (BOW). In
the limit of very strong on-site Coulomb interaction, the BOW instability is
usually referred to as the spin-Peierls (SP) instability. In the presence
of intersite Coulomb interactions, and for certain relative values of the
on-site and intersite interaction parameters, a charge-density wave (CDW),
periodic
modulation of the site charge density, can be the dominant 
instability \cite{Dixit}.
The BOW and the CDW occur in largely nonoverlapping regions of the parameter
space and compete against each other \cite{Dixit,Kivelson,Su,Hirsch}. 
True antiferromagnetism (AFM)---ie, a long-range order (LRO) 2k$_F$ 
spin-density wave (SDW)---is absent
in for spin-rotationally invariant models in 1D, and
the ground state is dominated by singlet spin coupling, which favors the
BOW over the SDW. Two-dimensionality is thus essential for the SDW.

The 1/2-filled isotropic
two-dimensional (2D) case has been investigated in great detail in recent years
(mostly for the case of large intrasite Hubbard interaction but zero intersite
interaction) 
\cite{Manousakis}, as this
limiting case is known to describe the parent semiconductor compounds
of copper-oxide based high temperature superconductors.
The BOW instability that characterises the 1D chain is destabilized in 2D
by Coulomb interaction \cite{Mazumdar,Prelovsek,Hirsch1}, and
the dominant broken symmetry here is the 
2k$_F$
SDW, with periodic modulation of the spin density. 
Most recently, it has been
demonstrated that this SDW state appears for the smallest nonzero interchain
hopping in weakly coupled 1/2-filled band chains \cite{Sandvik}, in agreement
with previous renormalization group calculations \cite{Affleck,Wang}.
As in 1D \cite{Hirsch}, there is no CDW-SDW coexistence in 2D
\cite{Mazumdar,Prelovsek,Hirsch1}.                                 
The absence of coexistence between the BOW and SDW for
the 1/2-filled band in both 1D and 2D can be readily
understood intuitively: the BOW requires
spin-singlet coupling between alternate nearest neighbor spins, which
clearly has to disappear in the SDW. An alternate way of viewing this is
to observe that the probability of charge-transfer to the left and to the
right in the AFM are exactly equal, and therefore the SDW cannot coexist
with the BOW. 
On the other hand, both the BOW and the SDW require that the
site-occupancies by electrons are strictly uniform, and thus neither the
1D BOW nor the 2D SDW will coexist with the CDW.

Coupled 1/2-filled band
chains have also been discussed within the context of the so-called
ladder systems \cite{ladder}. Whether or not a given n-leg ladder system,
for small n, exhibits the BOW now depends on whether n is odd or even. 
This feature of the ladder systems could have been 
anticipated from the physics of the 
odd versus even S Heisenberg chains
\cite{Haldane}.
Thus at least for the simplest monatomic
lattices, ground states of the 1/2-filled band
are known: the BOW, CDW and SDW phases
compete against one another and do not coexist, and 2D behavior emerges 
for the smallest 2D coupling.

In contrast to the 1/2-filled band, broken symmetries in
{\it non}-1/2-filled bands with strong e-e interactions
have been investigated primarily in
1D limit \cite{Solyom,Emery,Voit,Hirsch2,Mazumdar1}
or at most in the
quasi-1D regime of weak interchain coupling\cite{ladder-footnote}.
This emphasis likely arises from the theoretical preconception
that finite one-electron hopping between chains
destroys the nesting feature that
characterizes the 1D limit, leading necessarily to the restoration of the
metallic phase \cite{disorder}. A recent work has examined
coupled chains in the limit of weak
e-e interactions \cite{Fisher}. The weak-coupling approximation employed
in reference [23]
reproduces the loss of nesting predicted within band theory.
While the continuum renormalization group calculations
\cite{Solyom,Emery} predicted CDW-SDW coexistence for {\it incommensurate}
bandfillings, early quantum Monte Carlo calculations for the
1/4-filled band failed to find this coexistence \cite{Hirsch2}.
Many more recent
numerical simulations on discrete finite systems assume the
absence of coexistence between the 2k$_F$ BOW, the 2k$_F$  
CDW and the 2k$_F$ SDW that characterizes that 1/2-filled band also
applies to the non-1/2-filled bands.
Indeed, it is often assumed that the
CDW is driven by the e-ph interactions
and the SDW by e-e interactions
and that their effects are competing.
This assumption is made
despite the result mentioned above that already in the simplest case
of the 1D 1/2-filled band, e-e and e-ph interaction effects are known
not to be competing but to
act in a co-operative way to give the enhanced 2k$_F$ BOW \cite{review1}.   

Recently, we have begun a systematic study of the nature of the broken
symmetry ground states in the 2D
1/4-filled band on an anisotropic rectangular lattice
with both e-ph and e-e interactions\cite{mrcc,mcc}.
Earlier work by us had already established the
{\it cooperative} coexistence
between the BOW and the period 4 ``2k$_F$'' CDW in the 
\begin{figure}[htb]
\centerline{\epsfig{width=3.0in,file=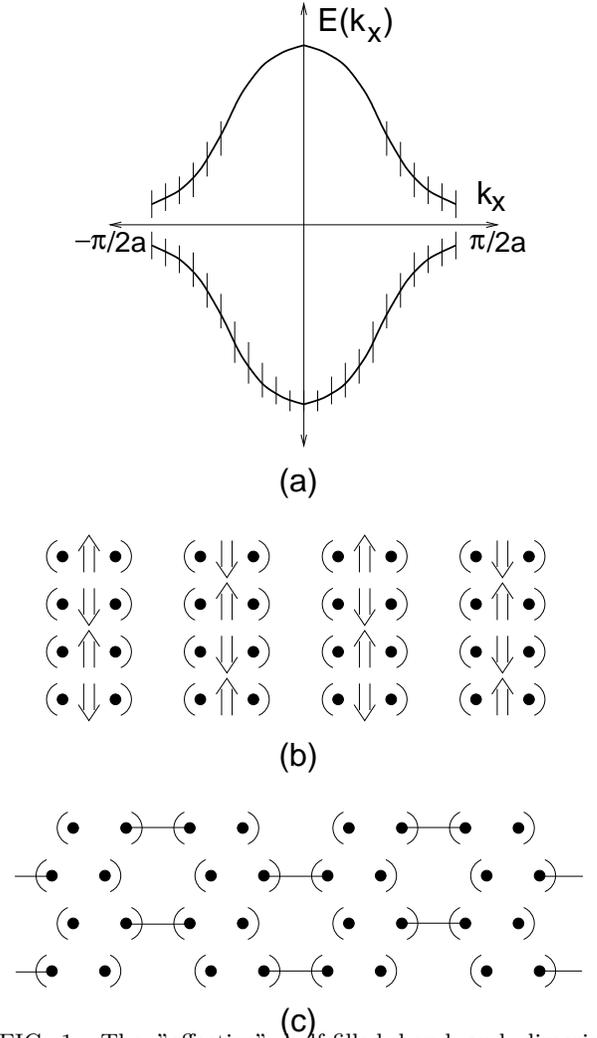}}
\caption{The "effective" half-filled band and dimerized dimer
model. (a) Dimerization in a 3/4-filled 1D band of {\it electrons}
leads to a gap in the single particle spectrum at $k= \pm\pi/2a$ ($a$ =
lattice spacing),
resulting in a half-filled upper subband. Note that although the actual
CTS materials are indeed nominally 3/4-filled electron
bands (hence 1/4-filled hole bands), in the text we follow 
the convention and refer to them simply as ``1/4-filled''. (b) A real space
depiction of a 2D lattice of dimers in the strong correlation limit.
The two sites within the parentheses form one lattice point of the
dimer lattice, and the intradimer bonds are stronger than the interdimer bonds.
The charge and spin populations on individual sites within each dimer are
equal, and the effective 1/2-filled band lattice is antiferromagnetic in 2D.
(c) Schematic of a frozen valence bond state resulting from the 
dimerization of dimer lattice.  
The interdimer bonds are now different; the line denotes a singlet bond.
This frozen valence bond diagram is relevant in the 1D limit, and then again
for the strongly 2D case,  where the antiferromagnetism has disappeared. The
antiferromagnetic phase that occurs for intermediate interchain coupling is
shown in Fig.~2.}
\label{eff_half}
\end{figure}
1D 1/4-filled band, with each broken symmetry enhancing the
other, for both noninteracting \cite{umc} and interacting \cite{umt}
electrons. The latter results have been subsequently confirmed by
 Riera and Poilblanc\cite{Poilblanc}.
In the more recent work \cite{mrcc,mcc}
we have demonstrated an apparently unique feature of the
1/4-filled band: namely, the coexistence of the BOW-CDW with the period 4 ``2k$_F$'' SDW,
giving rise to a coupled Bond-Charge-Spin density wave (BCSDW)
that appears for weak interchain electron transfer between chains. 

In the present paper, we extend our calculations
to the full range of anisotropies, from uncoupled chains
to an isotropic 2D lattice. We include both
the SSH intersite phonons that drive a BOW
\cite{review2} and
the Holstein phonons that drive a CDW \cite{Holstein}.
We list three primary motivations for this extension\cite{mcc}.
First, the cooperative coexistence between the BOW and the 2k$_F$ SDW
found in the 1/4-filled band for {\it weak} interchain transfer is exactly
opposite to the competition between the 2k$_F$ BOW and the 2k$_F$ SDW
(with the latter dominating for nonzero interchain transfer)
in the 1/2-filled band. It is then immediately natural to ask what the
nature of the ground state is for {\it strong} interchain hopping of
electrons in the 1/4-filled band.
Second, from a more general theoretical perspective,
whether or not the vanishing of density waves that is
predicted by one-electron nesting ideas remains true for strongly
correlated electrons is of considerable general interest. Finally, our
results are likely to have relevance to experimental observations
in the organic
CTS, including those that exhibit superconductivity 
\cite{Schulz,Ishiguro,Jerome}.

Our investigations yield the surprising result that
the coexisting Bond-Charge density wave (BCDW) persists as
the ground state of the strongly correlated 1/4-filled band in 2D for
{\it all} values of the
interchain electron transfer, including the isotropic limit.
We show that this result can be understood physically as
a consequence of
interchain confinement arising from strong intrachain
Coulomb interactions \cite{confinement1,confinement2,confinement3}.
The SDW
component of the BCSDW, on the other hand, attains a maximum amplitude at
some intermediate interchain transfer, after which it typically 
vanishes at a critical
value of the transfer.
 
In order to discuss applications of results to real materials, including the
2:1 cationic CTS, we need to
clarify an important aspect of our approach {\it vis-a-vis}
most previous work 
on models of these materials. In our above discussion
of band-filling, ``1/4-filled'' is defined in the usual manner:
namely, in the absence of the BCDW, the lattice is uniform in at 
least one direction,
and the average density of electrons {\it per site} is 1/2. 
In real materials, 
crystal structure effects often cause a lattice dimerization
that is unrelated to any underlying electronic or magnetic instability
(see below) \cite{Pouget1}. 
As shown in Fig.~\ref{eff_half}(a), this dimerization leads to a gap
in the single electron spectrum at $k_F=\pi/2a$, and consequently
suggests using an effective 1/2-filled band model that focuses
on the upper subband. In real space terms,
this approximation amounts to considering the system
as a set of (tightly bound) dimers ({\it i.e.}, a diatomic lattice) with
one electron per {\it dimer} site, as shown in
Figure.~\ref{eff_half}(b). This approach has been widely
applied\cite{Pouget1,Bourbonnais,Kanoda,Kino,McKenzie},
particularly with considerable success
in the context of the magnetic field-induced
spin density wave (FISDW) in 2:1 salts of TMTSF \cite{Chaikin,Gorkov}.  
As we show below, a further dimerization of the dimer lattice is unconditional
in both 1D (the well-known spin-Peierls transition) and 2D (a surprising
new result), and that this 
{\it dimerization of the dimer lattice leads
spontaneously to different electronic populations on the sites within a dimer,
i.e., to the same 
2k$_F$ CDW that occurs in the 1/4-filled (monoatomic) band} (see Fig.~1(c)). 
For small
interchain electron transfer, the BCSDW will therefore have nearly
the same structure
as the original 1/4-filled band. This is a third new result, perhaps also
surprising, and shows that the number of electrons per 
site within a unit cell is a more fundamental parameter
than the bandfilling: the latter is strictly a one-electron concept of limited
use in the interacting electron picture.

We expect our results to be relevant for the 1D semiconductors
(TMTTF)$_2$X, the
so-called ``quasi-1D''  organic superconductors (TMTSF)$_2$X,
as well as the 2D organic superconductors (BEDT-TTF)$_2$X and the more
recently synthesized (BETS)$_2$X \cite{nomenclature}.
In reference [24] we showed
that the highly unusual ``mixed CDW-SDW state''
\cite{Pouget1,Pouget2,Kagoshima}
found in (TMTTF)$_2$Br,
(TMTSF)$_2$PF$_6$ and (TMTSF)$_2$AsF$_6$
can be explained naturally as the BCSDW state
within the strongly correlated
1/4-filled band scenario. Our current work shows that dimerization of the
dimer lattice leads to the same results, and hence the weak
high temperature dimerization along the stack axis \cite{Pouget1}
is effectively irrelevant: starting from either the 1/4-filled model or the
effective 1/2-filled scenario, the final outcome is the same \cite{KO}.

With these comments complete, we can describe the organization of the 
remainder of the paper. In Section II we introduce our model Hamiltonian,
as well as that of the dimerized dimer model. In Section III, we
present physical, intuitive arguments, based on a configuration
space picture of broken symmetry\cite{Dixit,Mazumdar,Mazumdar1,umt} that
predict both the BCSDW for weak interchain
electron transfer and the persistent BCDW state in the isotropic limit.
In Section IV, we present the results of extensive numerical studies,
exploring behavior in both the strict 1D limit and for the full range
of anisotropies in the quasi-2D case. These studies, in confirmation
of the qualitative
predictions of Section III: (i) establish the persistence of the
BCDW up to the isotropic limit; (ii) suggest the occurrence of two quantum
critical transition as an SDW first appears for weak transverse
hopping and then disappears for the nearly isotropic case; and (iii) 
prove the equivalence of the 1/2-filled dimerized dimer and 1/4-filled
monatomic lattices. For clarity, in Section V we summarize our theoretical
conclusions; readers not interested in the underlying physical arguments
or numerical details can skip directly to this summary in Section V.
In Section VI, we examine in some 
detail several recent experiments that indicate the applicability
of our theory to the insulating states that are observed to be proximate
to the superconducting states in the organic CTS.
Finally, in Section VII, we indicate possible future directions
for our research, focusing on 
commensurability defects in the BCDW state and their
possible role in the proximate superconducting phases. We point
out several intriguing similarities between this potential
microscopic mechanism for superconductivity and other recent 
phenomenological models.
We conclude the article with three appendices, which deal with various more technical
arguments and details of the numerical methods.

\section{Models and Observables}

We consider two different extended Peierls-Hubbard Hamiltonians on a
rectangular lattice 2D with (in general) anisotropic electron hopping.
The first model describes a monatomic 1/4-filled band and is defined by the 
Hamiltonian
$$H = H_0 +H_{ee} + H_{inter} \eqno(1a)$$
$$H_0 = -\sum_{j,M,\sigma}[t-\alpha(\Delta_{j,M})]B_{j,j+1,M,M,\sigma}
+ \beta\sum_{j,M}v_{j,M}n_{j,M} $$
$$+ K_1/2\sum_{j,M}(\Delta_{j,M})^2 + K_2/2 \sum_{j,M}v_{j,M}^2 \eqno(1b)$$
$$H_{ee}=U\sum_{j,M}n_{j,M,\uparrow}n_{j,M,\downarrow} +
 V\sum_{j,M}n_{j,M}n_{j+1,M} \eqno(1c)$$
$$H_{inter} = 
-t_{\perp}\sum_{j,M,\sigma}
B_{j,j,M,M+1,\sigma} \eqno(1d)$$
\setcounter{equation}{1}

In the above, $j$ is a site index, $M$ is a chain index, $\sigma$ is
spin, and we assume a rectangular lattice \cite{mrcc,mcc,xysym}. 
As $t_{\perp}$ varies from 0 to $t$, the electronic
properties vary from 1D to 2D.
An implicit parameter in the above Hamiltonian
is the bandfilling, or more precisely $\rho$. We shall focus on the 
1/4-filled case, for which $\rho$ = 1/2. In applications to the organic CTS, 
each site is occupied by a single organic molecule, the displacement of
which from equilibrium is described by  $u_{j,M}$
(with $\Delta_{j,M}=(u_{j+1,M}-u_{j,M})$); $v_{j,M}$ is an intra-molecular
vibration, $n_{j,M,\sigma} = c_{j,M,\sigma}^\dagger c_{j,M,\sigma}$,
$n_{j,M} = \sum_{\sigma}n_{j,M,\sigma}$, and 
$B_{j,k,L,M,\sigma} \equiv
[c_{j,L,\sigma}^\dagger c_{k,M,\sigma} + h.c.]$, where 
$c_{j,L,\sigma}^\dagger$ is a Fermion operator. 
We treat the phonons in the adiabatic
approximation and are interested in unconditional broken symmetry solutions
that occur for e-ph
couplings $(\alpha, \beta) \rightarrow 0^+$. 
All energies such as $U$, $V$, and $t_\perp$ will be given in
units of the undistorted intra-chain hopping integral $t$.

The second model describes a {\it diatomic/dimer} lattice, with one electron 
per dimer.
The Hamiltonian for this case is similar to that above, with identical
$H_{ee}$ and $H_{inter}$, but with modified intrachain one-electron term
$H_0'$,

\begin{eqnarray}
H_0'& =& - t_1 \sum_{j,M,\sigma} B_{2j-1,2j,M,M,\sigma} \\
& &- \sum_{j,M,\sigma}[t_2 - \alpha \Delta_{j,M}]B_{2j,2j+1,M,M,\sigma} 
 + {K\over2}\sum_{j,M} (\Delta_{j,M})^2 \nonumber 
\end{eqnarray}

In the above each pair of sites (2j--1,M) and (2j,M) forms a dimer with 
fixed hopping
$t_1 > t$ between them, $\Delta_{j,M} = (u_{2j+1,M} - u_{2j,M})$, with
$u_{2j-1,M} = u_{2j,M}$; 
this means that there is no modulation of 
the intradimer bond
length, and the dimer unit is displaced as a whole. As written, the model
assumes an ``in-phase'' 2D
arrangement of the dimer units ({\it i.e.}, 
dimers on different chains lie directly
above one another), which we have determined to
be the lower energy configuration 
for both zero and nonzero $\Delta_{j,M}$. Notice that $H_0'$ does
not contain the Holstein on-site e-ph coupling. Nevertheless, we will show
that a site-diagonal CDW is a consequence of the BOW here.

The broken symmetries we are interested in are (i) the BOW, with periodic
modulations of the intrachain {\it nearest neighbor}
bond order $\langle\sum_{\sigma}B_{j,j+1,M,M,\sigma}\rangle$; 
(ii) the CDW,
with periodic modulations of the site charge-density
$\langle n_{j,M} \rangle$;
and (iii) the SDW, with periodic modulations of the site spin-density
$\langle n_{j,M,\uparrow} - n_{j,M,\downarrow}\rangle$. 
Note that in case of the dimer lattice (Eq.~(2)) we are interested in
both intra- and interdimer charge and spin modulations, although bond
modulations can occur only between dimers.
Furthermore, in the CDW and the
SDW the modulations of the site-based densities occur along both longitudinal
and transverse directions (though not necessarily with the same periodicities,
see below). In case of the BOW, a complete description would require the
determination of the phase difference between consecutive chains.

\section{Configuration space picture of spatial broken symmetry}
The physical arguments presented in this section provide crucial
insights that
allow us to anticipate the apparently counterintuitive results of this
paper. The need to develop such arguments arises from the limitations
inherent in all true many-body numerical simulations of strong
correlated electron systems: namely, one can study only systems of
limited size and distinguishing finite-size artifacts from true results
requires physical understanding. In turn, true many-body numerical methods
are essential here because of the intermediate
magnitude of the e-e interactions (comparable to the bandwidths) in the
organic CTS, which renders both mean field and
perturbation theoretic approaches questionable. For instance,
even in the strictly
1D limit, where well-established RG \cite{Solyom}
and bosonization\cite{Emery} techniques have existed for decades,
for the {\it intermediate} coupling regime,  
there have recently been some surprising discoveries
in the phase diagram of the extended
Hubbard model\cite{csc99,nakamura99}. 
In 2D, developing a clear physical intuition is still more crucial,
as numerically tractable lattices are even farther from the
thermodynamic limit, and the competition among broken symmetries
is likely to be more subtle.
Brief presentations of these physical ideas for $t_{\perp}$ = 0 \cite{umc,umt}
and
t$_{\perp} << t$ \cite{mrcc} have been made previously. Here we discuss 
these ideas 
for the complete range 0 $\leq$ t$_{\perp} \leq t$,
focusing on (i) the
transition from 1D to 2D, and (ii) the difference from the
1/2-filled band {\it monatomic} lattice.

A physical picture of spatial
broken symmetry in strongly correlated electron systems must necessarily be
based on configuration space ideas, as one-electron bands have simply ceased
to exist for strong e-e interaction.
Within the configuration space picture of broken symmetry
\cite{Dixit,Mazumdar,Mazumdar1},
each broken symmetry state,
independent of band-filling, can be associated with a small number of
equivalent configurations that are related by the symmetry operator in
question. For commensurate $\rho$, these configurations
are easily determined by inspection. The relevant configurations
consist of repeat units which
themselves possess the same periodicity as the density wave. For
illustration, we choose the 1D 1/2-filled band. In this case,
each broken symmetry has two extreme configurations, the pairs
corresponding to the SDW, BOW and CDW being, respectively: the two N\'eel states
${...\uparrow \downarrow \uparrow \downarrow...}$ and
${... \downarrow \uparrow \downarrow \uparrow...}$ (SDW);
the two
nearest neighbor valence bond diagrams (1,2)(3,4)(5,6)....(N -- 1,N) and 
(N,1)(2,3)(4,5)....(N -- 2, N -- 1) (where (i,j) is a spin singlet bond 
between sites i
and j and N is the number of sites) (BOW); and the 
configurations ...202020... and ...020202...(where the numbers denote
site occupancies) (CDW).
N applications of the one-electron hopping term in Eq.~(1)
on any one extreme configuration (corresponding to a given broken
symmetry) generates the other
extreme configuration, but for N ${\to \infty}$ this mixing of
configurations is small, and the ground state resembles one or the other
of the extreme configurations {\it qualitatively}, with reduced spin moment,
bond order or charge-density difference due to quantum fluctuations 
\cite{Dixit}.

The key insight of the configuration space heuristics
is that the qualitative effects of
many-body Coulomb interactions, as well as additional one-electron terms,
can be deduced from their effects on any one of the
extreme configurations \cite{Dixit,Mazumdar,Mazumdar1}. As a trivial
example of this,
a repulsive Hubbard $U$ destroys the CDW in the 1/2-filled band, 
simply because double occupancies 
in the extreme configuration ...202020...
``cost'' prohibitively high energy.
Significantly, in the 1/2-filled band, the 
extreme configurations favoring the SDW, the BOW and the CDW
are different, and there is a complete 
lack of overlap between
them. This essentially guarantees the absence of coexistence
among these broken symmetries in
both 1D and 2D.

To apply these ideas to the 1D 1/4-filled band,
we begin by considering 
the on-site charge configurations. A 2k$_F$ (4k$_F$) density wave here has 
period 4 (2) in configuration space. As discussed above, the extreme
configurations of interest must also have period 4 or 2, and there are
then only three distinct sets of extreme charge
configurations.
These contain the repeat units ...2000..., ...1100...,
and ...1010..., respectively, where the numbers again denote site occupancies.
There are four distinct configurations for sets 1
(...2000...) and 2 (....1100....), whereas there are only 2 for
set 3 (....1010....). By analogy with the 1/2-filled band (see above),
we now introduce spins and note that 
configurations belonging to sets 2 and 3 can again have spin singlet bonds
between pairs of nearest neighbor singly occupied sites, or the spins of the
occupied sites can alternate as in the 1/2-filled band N\'eel configurations.
Let us now show, by considering the different cases separately, 
how e-e interactions affect these configurations and how
an understanding of these effects suggests (correctly!) the
broken symmetries to be studied.

\subsection{1/4-filled band, $t_{\perp}$ = 0, $U = V$ = 0}

The non-interacting case provides a simple example to introduce some
of the important differences between the 1/4-filled and 1/2-filled bands.
Actual calculation indicates that within the 1D Holstein model the  
charge densities $\rho_j$ on the sites have the functional form \cite{umc}
\begin{equation}
\rho_j= 0.5+\rho_0 \cos(2k_Fja)=0.5+\rho_0\cos(\pi j/2)
\end{equation}
This charge density pattern could have been anticipated by focusing on the
extreme configuration ...2000..., which also predicts three different
charge densities (large, intermediate, small and intermediate),
since each `0' that is immediately next to a `2'
is different from the other pair of sites labeled `0' that are further away
from the `2'. 
occupancy scheme ...2000..., the probabilities of charge-transfer between
a `2' and the two neighboring `0's are
larger than that between the two neighboring `0's themselves. For 
nonzero $\alpha$ in Eq.~(1), this difference in charge-transfers leads to
lattice distortion of the form
\begin{equation}
u_j=u_0 \cos(2k_Fja)=u_0 \cos(\pi j/2),
\end{equation}
with bonding pattern ``SSWW'' (for strong, strong, weak,
weak), where a strong (weak) bond has hopping $t_S > t$ ($t_W < t$).
This then is one very important difference from the 1/2-filled 
band: whereas in the 1/2-filled band differences
in bond-orders arise
from spin-effects only (the probability of charge-transfer is greater
between nearest neighbor singlet-coupled sites than between nearest neighbor
non-bonded sites \cite{Dixit}), in non-1/2-filled bands this 
difference can also originate from site occupancies. 
{\it Precisely because the BOW and
the CDW here are both derived from the same extreme configuration,} they
coexist in the noninteracting 1/4-filled band \cite{umc}.

\subsection{1/4-filled band, $t_{\perp}$ = 0, $U, V > $ 0} 
For nonzero (positive) $U$ and $V$, the interplay among the various possible
broken symmetries becomes both more subtle and more interesting.
Since double occupancies ``cost'' energy, the extreme configuration
...2000... is suppressed even at a relatively small $U$ \cite{umt}.
For the strongly correlated
($U \rightarrow \infty$)
1D 1/4-filled band with convex long range interactions,
Hubbard showed that there exist
{\it two} different Wigner crystals, with occupancy schemes ...1100... and 
...1010...\cite{Hubbard}. 
At first glance, the extreme configuration ...1010..., corresponding to
a period 2 ``4k$_F$''CDW \cite{Hubbard}, 
appears to be strongly preferred, but
in fact more careful analysis shows that it dominates the
ground state only
for fairly substantial $V$ \cite{HF}.
This can be seen rigorously for ${U \to \infty}$,
where the 1/4-filled {\it spinful} band can be mapped rigorously
to the 1/2-filled {\it spinless} band \cite{Klein}, which in turn
can be mapped (via a Jordan-Wigner transformation) 
to an anisotropic Heisenberg spin 1/2 chain \cite{JW}. Using this
approach, one finds that the period 2 ``4k$_F$'' CDW becomes
the ground state only
for ${V > V_c = 2}$ (in units of $|t|$) \cite{exact}. 
For finite $U$, numerical
results\cite{Illinois} show that $V_c$ is
slightly larger than $2$. 
Given the estimated values of $V$ in
the organic CTS, it seems unlikely that they will exhibit this
(...1010...) intrachain ordering. This expectation is strongly supported
by the result that
the ...1010... CDW cannot coexist with the
BOW \cite{Dixit,Kivelson,Su,Hirsch}, whereas the (TMTTF)$_2$X are
known to exhibit a low-temperature transition
to a SP-BOW ground state \cite{Jerome}.

For ${V < V_c}$ the extended 1D Hubbard model at 1/4-filling
is a Luttinger liquid\cite{Haldane2} that is
also susceptible to a 2k$_F$ bond and charge distortion, and it is this
distortion that can be described by any one of the four equivalent
configurations ...1100...\cite{umt}. The 2k$_F$ CDW compatible with the
...1100... configuration has the form
\begin{eqnarray}
\rho_c(j)&=&0.5+\rho_0\cos(2k_Fja-3\pi/4) \nonumber \\
&=&0.5+\rho_0\cos(\pi j/2-3\pi/4),
\end{eqnarray}
This particular CDW also coexists with a BOW, since  
the charge-transfer across a `1 -- 1' bond
is different from that across a `1 -- 0' (or `0 -- 1') bond, 
which again is
different from the charge-transfer across a `0 -- 0' bond.
It is a subtle but crucial fact, confirmed by earlier numerical
studies\cite{umt}, that this same CDW can now promote {\it two}
different BOWs, each with three different bond strengths.
In each of these the `0 - 0' bond is the weakest, but depending upon the
strength of the Coulomb interaction,
the `1 - 1' bond can be stronger than a `1 -- 0' (or `0 -- 1') bond
(since charge-transfer in the former can 
occur in both directions), but it can also be weaker (since charge-transfer
in the former leads to double occupancy, while no double occupancy is created
in the charge transfer between a `1' and a `0'). Consistent with
this and the numerical results\cite{umt}, we shall
refer to the first bonding pattern as ``SUWU'' 
(for a strong `1 -- 1' bond, undistorted `1 --0' bond, 
weak `0 -- 0' bond, followed by an undistorted `0 -- 1' bond),
where a strong bond has $t_{S} > t$, an undistorted
bond has $t_{U}=t$, and a weak bond has $t_{W}<t$. This
BOW has pure period 4 ``2k$_F$'' periodicity and is accompanied by
lattice distortion 
\begin{equation}
u_j = u_0\cos(2k_Fja -\pi/4)=u_0\cos(\pi j/2-\pi/4).
\end{equation}
Again consistent with the numerical results, 
we call the second bonding pattern ``W$'$SWS''
(for a stronger weak `1 -- 1' bond, strong `1 -- 0' bond, weak `0 -- 0' bond
and strong `0 -- 1' bond, with $t_S > t >t_{W'} > t_W$).
Interestingly, the W$'$SWS bonding pattern is a superposition
of the pure 2k$_F$ period 4 SUWU structure and the pure 4k$_F$ period 2
SWSW structure and is accompanied by lattice distortion
\begin{eqnarray}
u_j &=& u_0[r_{2k_F}\cos(2k_Fja - \pi/4) + r_{4k_F}\cos(4k_Fja)] \\
\label{bond_mod}
    &=& u_0[r_{2k_F}\cos(\pi j/2 - \pi/4) + r_{4k_F}\cos(\pi j)] \nonumber, 
\end{eqnarray}
where $r_{2k_F}$
and $r_{4k_F}$ are the relative weights of the 2k$_F$ and 4k$_F$ bond
distortions, respectively \cite{umt}.
These results were established numerically in reference [27], 
where from comparisons
to available experimental data in the 1:2 anionic TCNQ systems it was
also shown that the
phase relationship between the coexisting 2k$_F$ CDW and the W$'$SWS BOW
(the W$'$ bond connects sites with greater charge densities than
the W bond) is precisely in agreement with theory. 

Very importantly, we show below that the 
dimerization of the dimer lattice with one electron per dimer also leads to a 
W$'$SWS
bonding pattern (see Fig.~\ref{eff_half}(c)), which in its turn promotes the 
site occupancy scheme  ...1100.... This coexistence will therefore occur
in either the full 1/4-filled band model or the effective 1/2-filled,
dimerized dimer approach.

\subsection{1/4-filled band, $t_{\perp} << t$, $U, V \neq$ 0.}

The above two BOW-CDWs describe the ground state of the interacting
1/4-filled band in the limit of $t_{\perp}$ = 0, where the `1 - 1' bond
is a singlet. As in the 1/2-filled
band though, singlets are expected to give way to SDW order for
${t_{\perp} \neq 0}$. Thus we must understand the role of the spin
degrees of freedom.
Once specific spins are assigned to the
sites labeled `1' in the ...1100... configuration, the 
sites labeled `0' become distinguishable, as a given `0' site is now
closer to one
particular `1' (up or down) than the other \cite{mrcc}. In this case the
`0' site is expected to acquire the spin characteristic of
its neighboring `1'. The charge and spin along a
chain can now thus be denoted as
$\uparrow,\downarrow$,{\tiny{$\downarrow,\uparrow$}}, where the sizes of
the arrows are schematic measures of the charge and spin densities on the
sites. Note that this represents the SDW of the form
\begin{eqnarray}
\rho_s(j) &\equiv& \langle c_{j,M,\uparrow}^{\dagger}c_{j,M,\uparrow} -
c_{j,M,\downarrow}^{\dagger}c_{j,M,\downarrow}\rangle \\
&=& \rho_{s2k_F} \cos (2k_F ja - \pi/4) + \rho_{s4k_F} \cos(4k_F ja - \pi), \nonumber
\end{eqnarray}
which {\it coexists} with the BOW and CDW.

Commensurability effects imply that the possible phase
shifts between adjacent chains in the anisotropic 2D system
are 0, $\pi$/2 and $\pi$, and we have performed
\begin{figure}[htb]
\centerline{\epsfig{width=3.0in,file=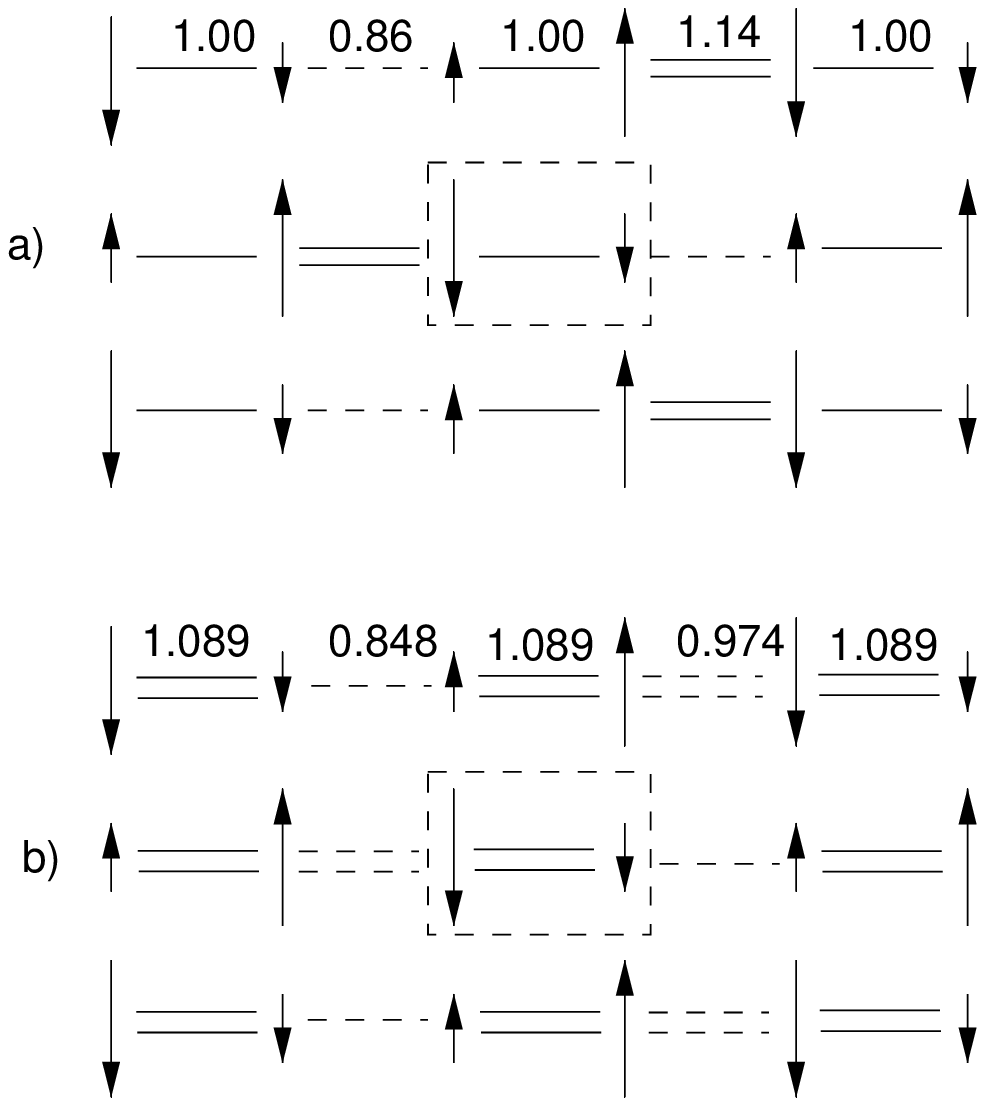}}
\medskip
\caption{Sketches of the BCSDW ground states that occur
for small $t_{\perp}$ in the
strongly correlated, anisotropic 2D 1/4-filled band.
The arrows indicate the spin directions and their
sizes indicate the relative charge and spin densities.
The hopping integrals used to calculate
the energies of the
distorted lattices correspond to (a) $r_{4k_F}$ = 0 (see text, Section V) and
(b) $r_{4k_F}$ = $r_{2k_F}$, and are shown above the bonds along the top chain.
This variation in $t$ reflects the BOW. The bond-distortion
pattern in
(b), with slightly modified weak bond hopping integrals, also corresponds
to the dimerized dimer lattice for small enough $t_{\perp}$.
Note that the charge ordering corresponds to the 1D paired
electron crystal along
the longitudinal {\it and} both diagonal directions and the the monatomic
Wigner crystal along the transverse direction.}
\label{cartoon}
\end{figure}
explicit numerical
calculations to determine that the lowest energy state is obtained with a
phase shift of $\pi$. The intrachain bond orders, 
determined by the probabilities of nearest neighbor charge-transfers, 
continue to be
different for the different pairs of neighboring sites. This is the
major difference between the possible broken symmetries in the 1/2-filled
and 1/4-filled band. While in the 1/2-filled band there is no overlap
between the extreme configurations favoring the BOW, CDW and SDW, in the
weakly 2D 1/4-filled band {\it the same extreme configuration supports all
three broken symmetries} \cite{mrcc}. For small nonzero $t_{\perp}$,
 we therefore
expect a strong cooperative coexistence between the BOW, the CDW and the
SDW. Furthermore, since the same CDW coexists with both the SUWU BOW and 
the W$'$SWS BOW, this coexistence is independent of which particular BOW
dominates. This has been explicitly demonstrated in
reference [24], where it was shown that the
overall ground state for small $t_{\perp}$ is one of the two BCSDW states
shown in Fig.~\ref{cartoon}, with overall 2D periodicity
of (2k$_F$, $\pi$).

\subsection{1/4-filled band, $t_{\perp} \leq t$, $U, V \neq$ 0.}

What happens as $t_{\perp}$ is further increased?
Within k-space single-particle theory, 
increasing $t_{\perp}$ should destroy
the nesting of the Fermi surface. But as we have indicated above, our real space
analysis predicts, and our numerical results will establish, that this
destruction does not occur. To argue this convincingly, we must first
show how this destruction of the nesting, which certainly does occur for non-interacting electrons,
can be correctly
described within our configuration space picture of the broken symmetry.
Recall that the one-electron hopping term in Eq.~(1) introduces ``paths''
between the extreme configurations, where each step in a given path connects
two configurations related by a single hop \cite{Dixit,Mazumdar,Mazumdar1}.
Nonzero $t_{\perp}$ introduces many additional
paths connecting the extreme configurations that are the 2D equivalents
of ...1100... (with a $\pi$-phase shift between consecutive chains). 
For $U = V$ = 0, there is no inhibition of these
paths, and it therefore becomes easier to
reach one extreme configuration from another, leading to  
enhanced configuration
mixing (relative to 1D), which in its turn destroys the ``nesting''
and the broken symmetry. 

The situation described above changes, however, for
nonzero Coulomb interaction. Interchain hopping $t_{\perp}$ leads to partial 
double occupancy on a single site
({\tiny $\uparrow$} $\downarrow$) with an energy barrier
that, while less than the bare $U$, is a $U_{eff}$ that increases with
$U$. The energy barrier to {\it inter}chain hopping leads to ``confinement''
of the electrons to single chains, a
concept that has been widely debated recently, in the context
of high $T_c$ superconductors\cite{confinement1,confinement2,confinement3}. 
For large enough $U_{eff}$, the confinement
can be strong enough that the broken symmetry state can persist
up to the isotropic limit $t_{\perp} \sim t$.

More precisely, the bond and charge components of the BCSDW 
can
persist up to the isotropic limit $t_{\perp} \sim t$, leading
to the BCDW state we have previously introduced.
The evolution of the spin structure is different from and
more subtle than the bond and
charge components. From
the cartoons in Fig.~\ref{cartoon}, we see that for the SDW to exist
it is essential that
the `0's have a spin ``direction''.
In the small $t_{\perp}$ case, the sign of the spin
on a `0' is
necessarily that of the nearest {\it intra}chain `1'.
Note, however, that each
`0' also has two {\it inter}chain `1's as neighbors and that
for a stable
SDW, the spin densities of the `1's that are neighbors of a specific `0'
must be opposite (as shown in the Figure).
Therefore, with increasing $t_{\perp}$, competing effects occur. On
the one hand, the
magnitude of the interchain exchange coupling $J_{\perp}$ $\sim
t_{\perp}^2/U_{eff}$ increases. On the other hand, the spin density
 on  a site labeled `0' decreases because of
the canceling effects of the {\it intra}- and {\it inter}-chain neighboring
 `1's. We thus expect the
SDW of the 2D lattice to vanish at a $t_{\perp}^c$ that will 
depend on the magnitudes
of the bare U and V.

This description of the evolution of the SDW applies to 
the true 1/4-filled band. In lattices that are dimerized initially,
further dimerization leads to the occupancy `10' 
\begin{figure}[htb]
\centerline{\epsfig{width=3.0in,file=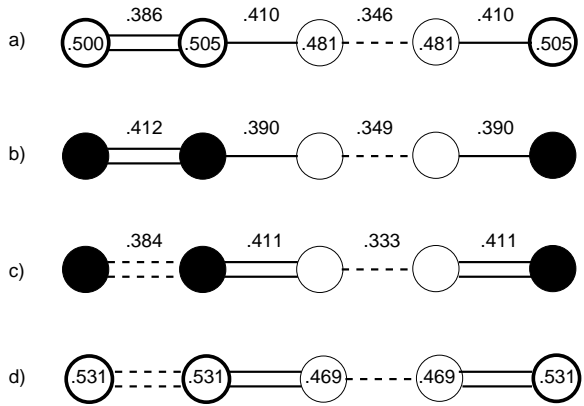}}
\medskip
\caption{Numerical results of 1D simulations for $U=6$, $V=1$. 
(a) Charge densities (numbers inside each circle, which represents one 
molecular site) and bond orders (numbers against the bonds)
at the center of an open uniform chain
of 16 sites for $\alpha = \beta = 0$.  
(b) Bond orders in a 16 site periodic ring with uniform hopping, and with
externally imposed 
period 4 magnetic field of the same form as in Fig.~\ref{cartoon},
with amplitude $\epsilon=0.05$. (c) Same as in (b) with $\epsilon$ = 0.1. 
Because of equal bond lengths and nonzero $V$, there is a weak contribution
by the ...1010... CDW to the ground state here and the charge densities are
not pure ...1100... The filled (unfilled) circles correspond to large (small)
charge densities. The bond orders also show weak deviation from pure
SUSU or W$'$SWS behavior, and the bond orders shown are averages for each kind 
of bond. The magnetic field induced SDW creates a spontaneous BCDW.
(d) Charge densities in a periodic dimerized dimer lattice of 16 sites.
The double bond corresponds to $t=1.2$, and the dotted and double dotted
bonds to $t=0.7$ and 0.9, respectively. Note that the CDW pattern in this
effective 1/2-filled band system is the same as the 1/4-filled band lattices
in (a), (b) and (c).}
\label{1d-fig}
\end{figure}
or `01' on each dimer. If
the original dimerization is very strong, the spin on a given `0' will 
continue to be strongly influenced by the spin on its partner in the dimer,
and $t_{\perp}^c$ at which the SDW vanishes in this case will be larger.

The robustness of the BCSDW and the BCDW relative to the
uniform metallic state can be understood from the cartoon occupancy schemes
in Fig.~\ref{cartoon}. 
It is instructive to discuss the BCDW state in terms of the two large $U$
Wigner crystal structures discussed by Hubbard\cite{Hubbard}.
We refer to the ...1100... electron arrangement
as that of a ``paired electron crystal'', and the ..1010... as the ``monatomic
Wigner crystal.'' For the
3D low density electron gas, Moulopoulos and
Ashcroft \cite{moulo} showed that there exists an intermediate density range 
where the paired
electron crystal has lower energy than the monatomic Wigner crystal,
and the region $0 < V < V_c$ in our discrete lattice case can be thought of as 
intermediate between the $V = 0$ and $V > V_c$.
A striking feature of the BCSDW and the BCDW occupancy
scheme is that it is a paired electron crystal along the chains (...1100..., 
periodicity 2k$_F$), a monatomic Wigner crystal 
transverse to the chains (...1010...., periodicity
4k$_F$), as well as a paired electron crystal along both diagonals
(...1100..., periodicity
2k$_F$). It is thus possible to predict that even in the presence of
interactions not explicitly
included in Eq.~(1), the BCDW continues to persist. For instance, by
enhancing the 4k$_F$ charge ordering along the transverse direction, the
nearest neighbor interchain Coulomb interaction $V_{\perp}$ will further
enhance the stability of the BCDW. Similarly, the diagonal ...1100... 
charge ordering implies that even the
additions of hopping $t_{diag}$ and Coulomb repulsion $V_{diag}$
along the diagonals
will not destroy the BCDW state
for realistic parameters: in particular, $V_{diag}$ stabilizes the
BCDW relative to the other Wigner crystal
(...1010...) along both $x$ and $y$ directions.

In the above our goal has been to predict a novel semiconducting state
that is more stable than the metallic state. Even if this semiconducting
state is assumed, however, there is an additional surprise in our claim,
viz., the dominance of the singlet BOW over the SDW for strong 
two-dimensionality in the interacting quarter-filled band. 
This is {\it exactly
opposite} to what is observed in the 1/2-filled band. While in the half-filled
band a single singlet-to-antiferromagnet transition occurs with increasing
$t_{\perp}$,  for the 1/4-filled band,
a second antiferromagnet-to-singlet transition is predicted
at large $t_{\perp}$. Since a full discussion of this second
transition at this junction would interrupt the flow of the
narrative, we defer it to Appendix 1, which presents
arguments based on variational concepts 
and valence bond theory to motivate this result. 

\section{Numerical results}

\subsection{Results for 1D lattices}

Computational limitations will compel us to use
fairly small lattices in 2D and will prevent us from studying
dynamical phonons (even at a classical, self-consistent level).
As a consequence, we will have to work with
explicitly distorted lattices, rather than allowing the distortions
to arise naturally, as they would in larger lattices calculated
with dynamical phonons. 
To provide justification for this approach, in this section we
(a) extend our previous 1D results obtained with nonzero $\alpha$ 
and $\beta$\cite{umt} to {\it zero} e-ph couplings, to demonstrate that
these bond and charge distortions are unconditional, and (b) show that
the dimerization of the
dimer lattice (see Eq.~(2)) leads to the same
CDW as the monatomic 1/4-filled band.

It is known that in a sufficiently long open chain the bond orders 
and the charge densities at the center of the chain
show the behavior in the long chain limit, even in the absence of the
e-ph coupling.
In Fig.~\ref{1d-fig}(a) we show the exact nearest neighbor bond orders and 
charge densities at the center of an open
{\it undistorted} chain of 16 atoms with all hopping integrals equal, 
for $U=6$, $V=1$. Note that
both the BOW and the CDW show the 2k$_F$ modulations discussed in section III,
and appear in spite of uniform hopping integrals.

Second, we recall that in a purely 1D system, a LRO SDW can occur
only if an external staggered magnetic field is applied.
We therefore incorporate an additional
(external field-like) term  
\begin{equation} 
H_{SDW}= -\sum_j\epsilon [n_{j,\uparrow} cos(2k_Fj)
+ n_{j,\downarrow} cos(2k_Fj+\pi/2)]
\label{field}
\end{equation}
and consider
${H + H_{SDW}}$ for the 1/4-filled band with amplitude $\epsilon=0.1$. 
In reference [27] the same Hamiltonian was investigated for the case
of finite bond distortion.
Figs.~\ref{1d-fig}(b) and (c) show the bond orders and CDW
for a periodic ring (zero e-ph coupling and {\it undistorted}
hopping integrals) with the SDW
$\uparrow \downarrow${\tiny $\downarrow \uparrow$} superimposed on it.
Note that because of the
periodicity, the bond orders are uniform for the finite ring for
$\epsilon$ = 0. For $\epsilon$ = 0.05 (Fig.~\ref{1d-fig}(b)) and
0.1 (Fig.~\ref{1d-fig}(c)), the externally imposed SDW creates 
{\it spontaneous} BOWs with r$_{4k_F}$ = 0 and r$_{4k_F} \neq$ 0, respectively.

In Fig.~\ref{1d-fig}(d) we show the charge densities on a {\it periodic} ring
of 16 sites, now for the dimerized dimer lattice (the hopping integrals
here are 1.2, 0.9, 1.2
and 0.7). The charge modulations (which appear entirely due to modulations
of the {\it interdimer} bond orders) on the sites are exactly as in
Figs.~\ref{1d-fig}(a)--(c),
with the larger charges occurring on the sites 
connected by the stronger weak
bond (the W$'$ bond, with $t_{W'}$ = 0.9). In discussions of the spin-Peierls
transition within the effective 1/2-filled band (corresponding to the
dimer lattice), it is usually assumed that the
electronic populations within each dimer cell remains uniform in the
spin-Peierls state. Fig.~\ref{1d-fig}(d) clearly shows that this is not true.

\subsection{Results for 2D lattices}

To confirm the expectations based on the qualitative arguments
of Section III, we use
exact diagonalization and Constrained Path
quantum Monte Carlo (CPMC) \cite{CPMC} numerical techniques 
to calculate for representative finite 2D lattices: (i) the electronic energy
gained upon bond distortion,
\begin{equation}
\Delta E \equiv E(0) - E(u_{j,M}), 
\end{equation}
where $E(u_{j,M})$ is the electronic
energy per site with {\it fixed}
distortion $u_{j,M}$ along the chains; 
(ii) the site
charge densities $\rho_{j,M}$ for the bond-distorted lattices; 
due to the coexistence
of the BOW and the CDW, measuring the  CDW amplitude that results as a
consequence of the external modulation of the hopping integrals
is exactly equivalent to the measurement of the
bond order differences in the charge-modulated lattices;
and (iii) the z-z component
of the spin-spin correlations, for a range of $U$, $V$ and $t_{\perp}$.
We consider three distinct distorted lattices, 
two of which correspond to those shown
in Figs.~\ref{cartoon}(a) and (b), where we have indicated the 
hopping integrals along
the chain (the uniform lattice has a hopping integral of 1.0 corresponding
to all intrachain bonds). The third distorted lattice we consider is the
dimerized dimer lattice, the hopping integrals for which 
will be discussed later.

Ideally, calculations that aim to demonstrate persistence of a spatial
broken symmetry should do fully self-consistent calculations of the total
energy, which is a sum of the the electronic energy gain $\Delta E$
(including
effects of both e-e and e-ph interactions) and
the loss in lattice distortion energy. Unfortunately, in true many-body
simulations (such as exact diagonalizations or CPMC) of the very large
2D lattices we investigate (see below), such self-consistent
calculations are not possible. A well-tested alternate approach
\cite{review1} is to
calculate only the electronic energy gain for {\it fixed} lattice distortion
and compare the calculated $\Delta E$ against a known reference configuration,
where the distortion is {\it known} to occur.
This approach works because for a fixed distortion, the contribution of the
elastic energy to the total energy is constant, independent of the
other parameters; therefore the gain in electronic energy, relative to
that for the reference configuration, is a direct measure of the tendency to
distortion. An example of a previous successful application of this
approach is the enhancement by e-e interactions of the bond alternation in
the 1D 1/2-filled band; here, the reference configuration corresponds to the
limit of zero e-e interaction (SSH model),
where the Peierls bond alternation is known to occur \cite{review2}.  For
nonzero e-e interaction, the electronic energy gain for fixed bond
alternation can be larger (see Figs. 2.26 and 2.31 in reference
\onlinecite{review1}), indicating the enhancement of the bond alternation
by e-e interaction, a theoretical result that has been confirmed by all
subsequent studies. Similarly, in the 2D 1/2-filled band, calculations of the
electronic energy gain for fixed bond distortion have been used to prove the {\it
decrease} in the tendency to Peierls bond alternation upon the inclusion of
e-e interaction 
(see Fig.~10 in reference \onlinecite{Hirsch1}), a result that is in agreement
with other studies \cite{Mazumdar,Prelovsek} as well as the determination of
long range AFM in this case \cite{Manousakis}. Thus the approach has been shown
to work in two cases in which exactly opposite outcomes, -- in one case, an
increase in dimerization, in the other case, a decrease, occurred, indicating its
robustness.

At first glance, it appears that there exist two different reference
configurations in the present case. 
First, for given
$t_{\perp}$, one could study $\Delta E$ as a function of $U$ and
$V$: in essence, this amounts to comparing {\it uncorrelated} and 
{\it correlated} lattices for each $t_{\perp}$. Second,
for given $U$ and $V$, one could calculate $\Delta E$ as a
function of $t_{\perp}$. In fact, the first approach does {\it not}
yield correct results for two reasons: (i) the uncorrelated
2D lattices are undistorted, so there is no obvious $\Delta E$ with
which to compare the correlated results; and (ii) magnitude 
of $\Delta E$ {\it decreases} with $U$ and $V$ even
in the 1D limit, where we {\it know} that the bond and charge 
distortions are unconditional (see references
\onlinecite{Solyom,Emery,Voit,Hirsch2,umt}, as well as
the immediately previous subsection on 1D numerical results).
Thus to determine properly the tendency to distortion
in 2D, our reference configuration should be the single chain. We therefore
normalize the energy gained for coupled chains ($\Delta E$) against that for 
the single
chain ($\Delta E_0$) with the same $U$ and $V$.
A decreasing $\Delta E/\Delta E_0$ as
a function of $t_{\perp}$ signals the destruction of the
distortion by increasing two-dimensionality, while a constant or increasing
$\Delta E/\Delta E_0$ indicates a persistent
distortion \cite{review1,Dixit}.
Since the BOW and the CDW are coupled
cooperatively, the behavior of the charge ordering gives a second measure
for the tendency to bond distortion. Decreasing charge ordering for {\it fixed} bond distortion, as a function of $t_{\perp}$ (as occurs for noninteracting
electrons),
indicates the tendency to decreasing bond distortion, 
while constant or increasing
charge ordering indicates persistent bond distortion. The expected (and 
calculated, see below) charge ordering pattern is the same for all bond
distortion patterns and is the same as in 1D (with, however, a $\pi$-phase
shift between consecutive chains).

As mentioned above, our numerical calculations involve both exact 
diagonalization and the CPMC technique. Because of the sign errors
that plague quantum Monte Carlo calculations in 2D, it is critical to obtain
a precise idea about the accuracy of the numerical results. This is
especially so because CPMC calculations that have been reported so far
\cite{CPMC,Guerrero}
are only for the simple Hubbard Hamiltonian and did not include the nearest
neighbor interaction $V$. In Appendix 2 we discuss our
methodology and give detailed comparisons of energies and
correlation functions obtained for finite lattices within the CPMC and
exact diagonalization procedures. As shown there, although the CPMC technique
is not variational, the accuracies in both energy and correlation functions
are sufficient for our purposes.

\begin{figure}[htb]
\centerline{\epsfig{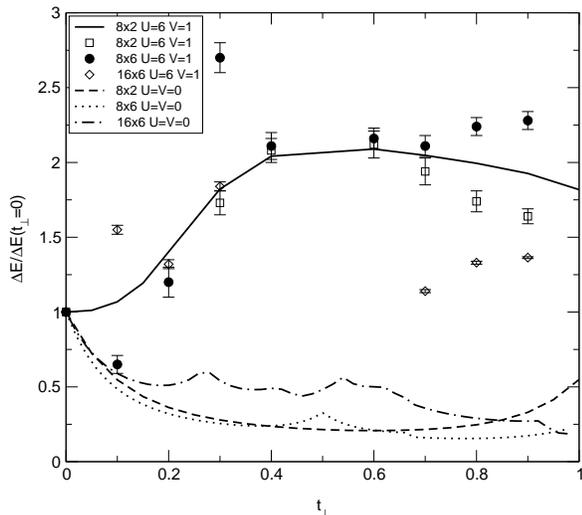}}
\caption{$\Delta E/\Delta E_0$ versus $t_\perp$ for a 2k$_F$
bond distortion ($r_{4k_F}=0$)
for the 8 $\times$ 2, 8 $\times$ 6, and
16 $\times$ 6 lattices for $U = V = 0$ and for $U = 6$, $V = 1$. For the
8 $\times$ 2 lattice both exact (solid line) and CPMC results are shown.
Intrachain hopping integrals for the distorted lattices
are as indicated in Fig.~\ref{cartoon}(a).}
\label{delta_e_2kf}
\end{figure}
For numerical results obtained from finite-size calculations to be relevant
in the thermodynamic limit, it is essential to choose
proper boundary conditions.
In the present case, we choose lattices and boundary 
conditions based on the physical requirement that {\it for noninteracting
electrons any nonzero
$t_{\perp}$ must destabilize the BCDW on that 
particular finite lattice}. Details of the analysis 
that guided our choice of
2D lattices are also presented in Appendix 2.
There we show N $\times$ M lattices 
\begin{figure}[htb]
\centerline{\epsfig{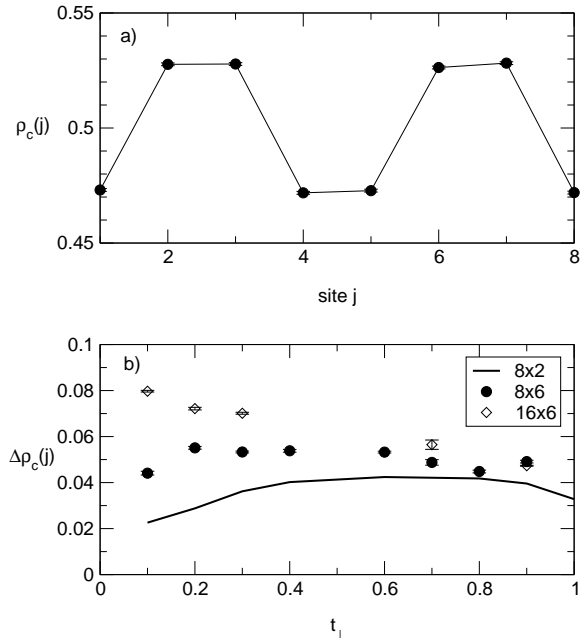}}
\caption{(a): Site charge densities on
one of the 6 chains in a 2k$_F$ bond-distorted 8 $\times$ 6 lattice, for
$t_\perp=0.2$, $U=6$, and $V=1$. The line is meant as a guide to the eye.
Note the expected "..1100.." structure discussed in the text.
(b) Amplitude of the 2k$_F$ CDW for the 2k$_F$ bond-distorted
8$\times$2 (exact), 8$\times$6, and 16$\times$6 lattices.
The ground state of the 16$\times$6 lattice is in the S = 1 subspace
for $t_\perp > 0.6$, and the CDW amplitudes for the S = 0 states here
are expected to be greater than those calculated for the ground state
and shown in the Fig. (see text).}
\label{delta_rho}
\end{figure}
(with N the number of sites per chain
and M the number of chains) that obey the above physical
requirement are restricted to those for which N = 8n, where n is an integer.
On the other hand, there is no restriction on M, except that M be
even to avoid even/odd effects. In our calculations below, we have chosen
M = 4n + 2, for reasons that are also discussed in Appendix 2.

We make one final point before presenting the 2D numerical data.
The restriction to N = 8n sites coupled with the 1/4-filling
introduces a potential subtlety into the
numerical computations of $\Delta E/\Delta E_0$ for nonzero $U$ and $V$.
Finite 4n-electron non-1/2-filled 1D undistorted periodic
rings have their ground state in the total spin S = 1 subspace, and even
the distorted system's ground state can be in the S = 1 subspace for the
smallest 4n-electron rings. We have confirmed from
exact
\begin{figure}[htb]
\twocolumn[\hsize\textwidth\columnwidth\hsize\csname @twocolumnfalse\endcsname
\centerline{\epsfig{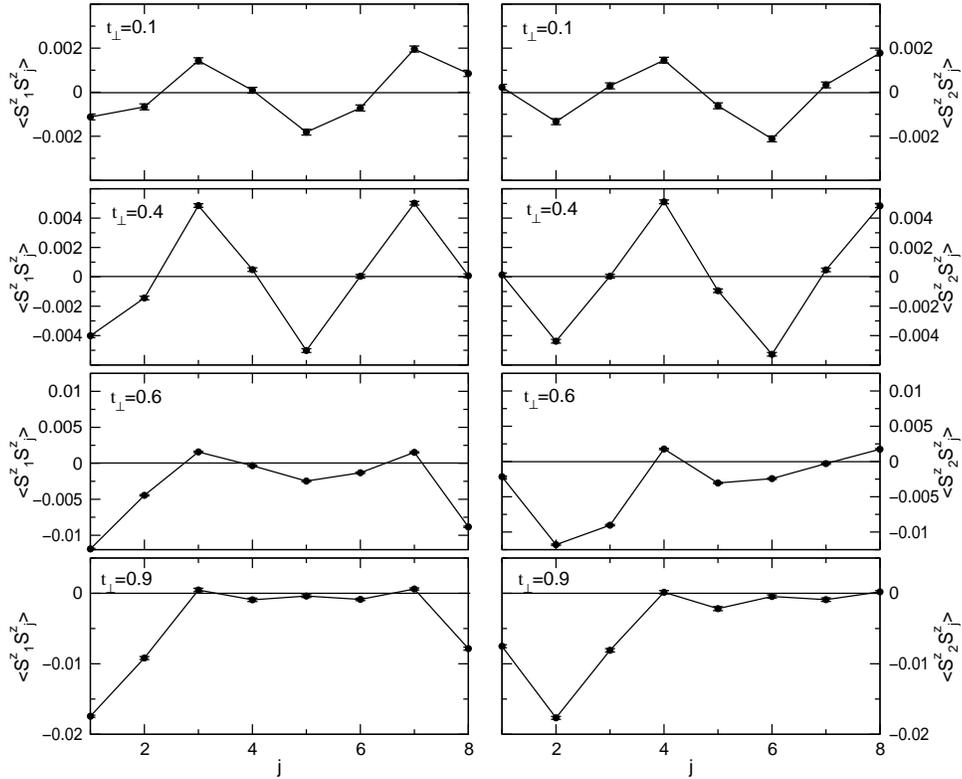}}
\caption{The z-z spin correlations between sites 1 (left panels) and
2 (right panels) on
the first chain of the 8 $\times$ 6 lattice and sites j = 1 -- 8 on the second
chain, with $U = 6$, $V = 1$ for four values of $t_{\perp}$.
Due to finite
size effects the wavefunction has small admixing with the ...1010... charge
order which affects the individual magnitudes of the
spin-spin correlations (see text).
AFM correlations increase with $t_{\perp}$ up to $t_{\perp} = 0.4$ but then
vanish at $t_{\perp} \simeq 0.6$, even though the BCDW continues to persist
for all $t_{\perp}$ (see Fig.~\ref{delta_rho}).
Lines are guides to the eye.}
\label{8x6_spin_spin}]
\medskip
\end{figure}
 diagonalizations of the 8 $\times$ 2 lattice that the ground state is
in the S = 0 state for the smallest nonzero $t_{\perp}$.
Thus while $\Delta E_0$
can correspond to the
energy gained upon distortion in the S = 1 subspace, $\Delta E$ necessarily
corresponds to the energy gained upon distortion in the S = 0
subspace. As this important but subtle point requires extensive
discussion that would interrupt the presentation here,
we present the details in Appendix 3, where we show
that despite this subtlety, the behavior of $\Delta E/\Delta E_0$
nevertheless is a proper measure of the stability of the distorted state
for nonzero $t_{\perp}$.

\subsubsection{Exact diagonalization and CPMC calculations, r$_{4k_F}$ = 0}

In Fig.~\ref{delta_e_2kf}
we show the behavior of $\Delta E/\Delta E_0$
for the non-interacting and interacting ($U = 6$, $V = 1$) cases
for three different lattices satisfying our boundary condition constraints.
In all cases we measure the electronic energy gained upon 
2k$_F$ SUWU bond distortion
(corresponding to nearest neighbor hopping integrals
$t_S$ = 1.14, $t_U$ = 1.0,
and $t_W$ = 0.86), relative to that of the undistorted state with equal
hopping integrals.
For the 8$\times$2 lattice the calculations involved both exact
diagonalization and the CPMC technique. 
The 8$\times$2 results, taken together,
then provide an estimate of the precision of the CPMC calculation.
The exact diagonalization
studies also confirm that the system is in the total spin state S = 0 for
$t_{\perp}$ as small as 0.01 (see Appendix 3).

The large scatter in the normalized $\Delta E$
at very large and very small $t_\perp$ may be due to the
degeneracies in the
non-interacting system at $t_\perp\rightarrow 0$ and $t_\perp\rightarrow 1$.
Furthermore, as pointed out in Appendix 2 (subsection A),
the absolute values of $\Delta E$ are
rather small, especially for the pure 2k$_F$ (r$_{4k_F}$ = 0) distortion.
The systematic errors due to the CPMC approximation are therefore large
in these two regions.
Nevertheless, except for the $\Delta E/\Delta E_0$ value at $t_\perp=0.1$
for the 8$\times$6 lattice, at all other $t_{\perp}$
the $\Delta E/\Delta E_0$ values are above 1 for all three lattices, and far
above the normalized non-interacting values.
As seen in Fig.~\ref{delta_e_2kf},
while for the non-interacting cases the 
$\Delta E/\Delta E_0$
decreases rapidly with $t_{\perp}$, for the interacting cases
the $\Delta E/\Delta E_0$ either remains unchanged
or is enhanced by $t_{\perp}$. Because of the strong degeneracies in the
one-electron occupancy scheme at the Fermi level at $t_{\perp}$ = 1, a
single well-defined one-electron wavefunction is missing here. The CPMC
calculations therefore could not be done for $t_{\perp}$ = 1.0. It is,
however, highly unlikely that the BCDW persists for $t_{\perp} = 0.9$ but
vanishes at $t_{\perp} = 1$; this expectation is corroborated
by the results of the exact diagonalization
studies for the 8 $\times$ 2 lattice, which were performed for the full range
of $t_{\perp}$, including $t_{\perp} = 1$ and showed enhanced distortion
throughout the whole region.
In the following sections we also show $\Delta E/\Delta E_0$ for the
\begin{figure}[htb]
\twocolumn[\hsize\textwidth\columnwidth\hsize\csname @twocolumnfalse\endcsname
\centerline{\epsfig{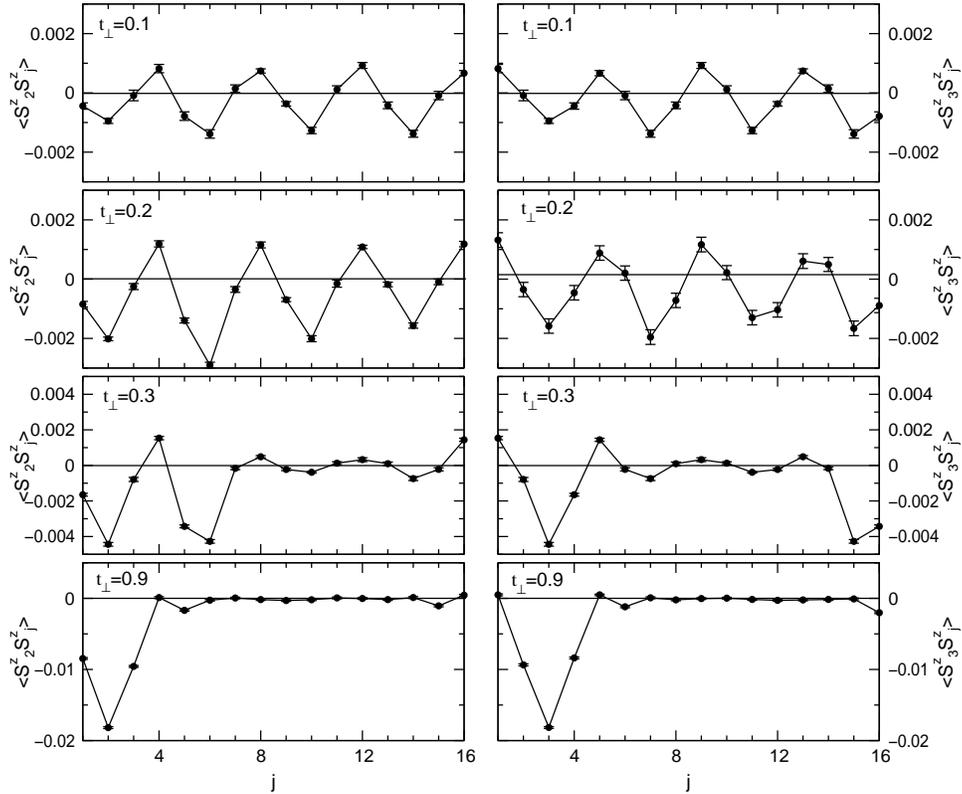}}
\caption{The z-z spin correlations between sites 2
(left panels) and 3 (right panels)
on the first chain of the 16 $\times$ 6 lattice and sites j = 1 -16
on the second
chain, with $U = 6$, $V = 1$ for four different values of $t_{\perp}$.
The finite size effects, and contamination with the ...1010... charge order
here is smaller than in Fig.~\ref{8x6_spin_spin}.
Lines are guides to the eye.}
\label{16x6_spin_spin}]
\medskip
\end{figure}
2k$_F$+4k$_F$ (r$_{4k_F}$ $\neq$ 0) and dimerized dimer
lattice. In both of these cases, the magnitude of $\Delta E$
is larger and hence easier to compute, but degeneracies restrict
CPMC simulations to smaller $t_\perp$. In both cases,
$\Delta E/\Delta E_0$ is close to or above 1 for all $t_\perp$
we have studied.

As discussed in the above, the bond-distorted lattices (both $r_{4k_F}$ = 0
and $r_{4k_F} \neq$ 0) have a synergetic coexistence with the CDW.
Thus the amplitude of the CDW, defined as $\Delta \rho_c = 
\rho_{cl} - \rho_{cs}$,
where $\rho_{cl}$ and $\rho_{cs}$ are the larger and smaller 
charge densities on the
...1100... 2k$_F$ CDW, is an alternate
measure of the stability of the BOW. If the nonzero $t_{\perp}$ destabilized
the bond-distortion, then even with {\it fixed} 2k$_F$ distorted hopping
integrals the amplitude of the BOW (measured as the differences in the
bond orders) would decrease, and the diminished strength of the
BOW in turn would decrease $\Delta \rho_c$. This is easily confirmed for the
noninteracting Hamiltonian, where
the amplitude of the CDW
decreases with increasing $t_{\perp}$.
In Fig.~\ref{delta_rho}(a) 
we show the charge densities on a single chain
for a bond-distorted 8$\times$6 lattice (because of periodicity,
all chains are equivalent)
for $U = 6$, $V = 1$, and $t_{\perp} = 0.2$. 
In Fig.~\ref{delta_rho}(b) 
we have shown the behavior of $\Delta \rho_c$ for all
the three lattices we have studied, now as a function of $t_{\perp}$.
Degeneracies in the one-electron energy levels in the 16$\times$6 lattice
for $t_{\perp} > 0.6$ even with finite bond-distortion cause
the CPMC ground states 
in this region to be S = 1.
Exact calculations in the 1D limit show that the amplitude
of the CDW in S = 1 is less than that in S = 0. Thus the weak 
decrease in the
$\Delta \rho_c$ values with $t_{\perp}$ in the 16$\times$6 lattice is a spin
effect: the bond distorted state is S = 0 at small $t_{\perp}$ and S = 1
at large $t_{\perp}$. The $\Delta \rho_c$ values 
at large $t_{\perp}$ for the 16 $\times$ 6 lattice
should therefore be considered as {\it lower
limits} (the $\Delta \rho_c$ values of the 16 $\times$ 6 lattice
are considerably larger than that of the S = 1
single chain of 16 sites).
In agreement with the behavior of the
$\Delta E$ in the interacting case (see Fig.~\ref{delta_e_2kf}), 
the CDW amplitude now {\it increases}
or remains constant with increasing $t_{\perp}$ for all the lattices studied,
indicating a greater tendency to bond and charge
distortion with increasing $t_{\perp}$. {\it Taken together, the results of
Figs.~\ref{delta_e_2kf} and 
provide quantitative proof of our qualitative
arguments establishing that the BCDW is a robust broken symmetry 
state for the interacting 2D ${1\over 4}$-filled band.}

In Fig.~\ref{8x6_spin_spin} we show
the inter-chain spin-spin correlations between sites 1 and 2 on the
first chain, and sites j = 1 -- 8 on the second chain, for the 2k$_F$
bond-distorted 8$\times$6 lattice for several values of $t_{\perp}$. 
The SDW profile is somewhat different from what is expected from a pure
...1100... charge modulation along the chains because the wavefunction of 
this finite lattice
also has contributions from the ...1010... type 
intrachain charge modulation.
The small ...1010... contribution to the wavefunction affects
the charge density, $n_{j,M,\uparrow} + n_{j,M,\downarrow}$ only weakly,
but the spin density, being the difference 
$n_{j,M,\uparrow} - n_{j,M,\downarrow}$ is a smaller quantity and is
affected relatively
more strongly. It is useful here to recall however that within the rectangular
lattice, ...1010... charge orderings along both longitudinal and transverse
directions give triangular lattice of occupied sites, and thus a pure 
...1010...
cannot give the SDW profiles of Fig.~\ref{8x6_spin_spin} (see
also below) \cite{mrcc1}.

Qualitatively, at
$t_{\perp}=0.1$ the SDW behavior is the same as in Ref.~\onlinecite{mrcc},
where these calculations were done for the 12$\times$4 lattice: the amplitude
of the interchain spin-spin correlation is independent of the distance 
between the sites, indicating long-range order. 
The qualitative
behavior of the spin-spin correlations is the same for  
$t_{\perp}$ = 0.4, where, however, the amplitude of the SDW is larger.
At still larger $t_{\perp} (= $ 0.6), the inter-chain correlations are very
strongly antiferromagnetic at short distances
(j = 1,2 on chain 2),
but the antiferromagnetic
correlations have disappeared at larger distances. This can be seen from
comparisons of the spin-spin correlations corresponding to values of j 
lying near the center of the second chain (j = 5), which are farthest from the
spins occupying sites 1 and 2 on the first chain. 
While the spin-spin correlations
near j=5 increase from $t_{\perp}$ = 0.1 to 0.4, they decrease
as $t_{\perp}$ is further increased to 0.6.
Similarly, focusing on site 8 of the second chain, we see that the spin-spin
correlation with site 1 on the first chain has actually changed sign upon
increasing $t_{\perp}$ to 0.6 from 0.4 (due to
the very strong short-range antiferromagnetic correlations), and the
magnitude of the positive spin-spin correlation with site 2 on the first chain
has decreased. All of these results indicate the
absence of long-range spin order for large $t_{\perp} \geq$ 0.6 in the 
8$\times$6 lattice. The loss of the long-range spin-order is most clear at
$t_{\perp}$ = 0.9, where spin-spin correlations are nonzero only for the
nearest interchain neighbors.

Fig.~\ref{16x6_spin_spin} shows the inter-chain spin-spin
correlations between sites 2 and 3 on the first chain 
and sites j=1$\ldots$16 on the second chain for the 16$\times$6 lattice.
The admixture of the intrachain ...1010... CDW is weaker in this larger
system: this is because
the ``tunneling'` between the extreme configurations ...1100... and, 
say, ...0110..., decreases with size, and as consequence, $V_c$ increases
with size in finite systems.
This can be seen by simply comparing the figures on the left
and right panels for $t_{\perp}$ = 0.1 and 0.2. If the intrachain CDW were
a pure ...1010..., 
\begin{figure}[htb]
\centerline{\epsfig{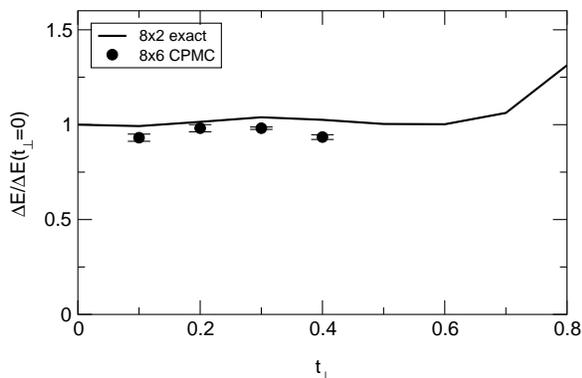}}
\caption{$\Delta E/\Delta E_0$ versus $t_{\perp}$ for
$r_{4k_F}$ = $r_{2k_F}$ for the 8 $\times$ 2 and 8 $\times$ 6 lattices
for $U = 6$, $V = 1$.
Intra-chain hopping integrals for the distorted lattices
are as indicated in Fig.~\ref{cartoon}(b).}
\label{delta_e_2kf4kf}
\end{figure}
the signs of the spin-spin correlations for each j would
be the same for both i = 2 and i = 3. Different signs for these correlations
are signatures of the ...1100... CDW (see Fig.~\ref{cartoon}).
As in the 8$\times$6 system, long-range SDW behavior is seen for 
$t_{\perp} = 0.1$. 
Focusing on sites j=7 -- 12 on the second chain, 
the amplitude of the SDW increases from $t_{\perp} = 0.1$ to
$t_{\perp} = 0.2$,
but further increasing $t_{\perp}$ to 0.3 
destroys the long-range order, as evidenced again 
by very large
AFM correlations at short distances and vanishing correlations
at large distances (sites j=7$\ldots$12 on the second chain).
The vanishing of the SDW is seen most clearly at
very large $t_{\perp}$ ($t_\perp=0.9$ in Fig.~\ref{16x6_spin_spin}).
We observe this same behavior of the
SDW on 8 $\times$ 2 lattice. In all cases, the SDW amplitude initially 
increases, exhibits a maximum,  and then vanishes
at a $t_{\perp}^c$ which decreases with the size of the system. As discussed
in section III.D, this behavior is to be expected from the nature of the
BCSDW in Fig.~\ref{cartoon}. 
The initial increase of
the SDW amplitude indicates that $t_{\perp}^c$ is nonzero, a
conclusion that is also in
agreement with the experimental observation of the BCSDW state in
the weakly 2D organic
CTS (see below). Based on the calculations for 16 $\times$ 6
lattice, we estimate
$0.1 < t_{\perp}^c < 0.3$ for the strictly rectangular lattice for
$U = 6$, $V = 1$. 

\subsubsection{Persistent distortions with r$_{4k_F}$ $\neq$ 0} 

The bond modulation pattern in the 1/4-filled band
given in Eq.~(7)
has in general both
2k$_F$ and 4$k_F$ components.
Figs.~\ref{delta_e_2kf} and \ref{delta_rho} 
show persistent distortion at large inter-chain
couplings for $r_{4k_F}$ = 0 (purely 2k$_F$ bond distortion). 
The persistent BCDW is expected also for $r_{4k_F} \neq$ 0.
\begin{figure}[htb]
\centerline{\epsfig{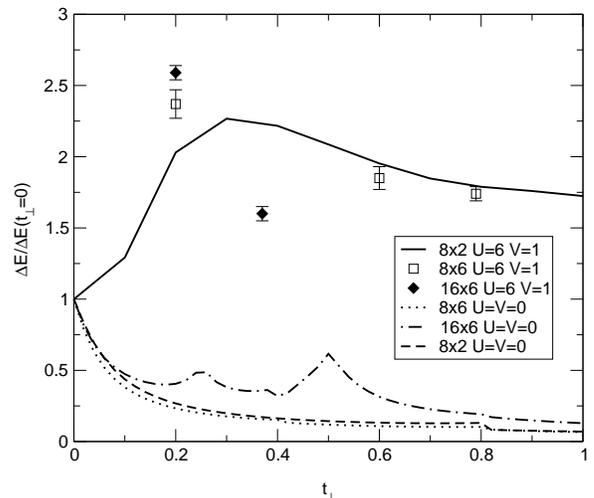}}
\caption{$\Delta$E/$\Delta$E$_0$ versus $t_{\perp}$ for
a dimerized dimer lattice
for the 8$\times$2, 8$\times$6 and 16$\times$6 lattices,
for $U = 6$, $V = 1$. The intradimer hopping integrals are 1.2 in both cases.
All interdimer hopping integrals are 0.8 in the dimer lattice,
and 0.7 and 0.9 in the dimerized dimer lattice.}
\label{delta_e_dd}
\end{figure}
\begin{figure}[htb]
\twocolumn[\hsize\textwidth\columnwidth\hsize\csname @twocolumnfalse\endcsname
\centerline{\epsfig{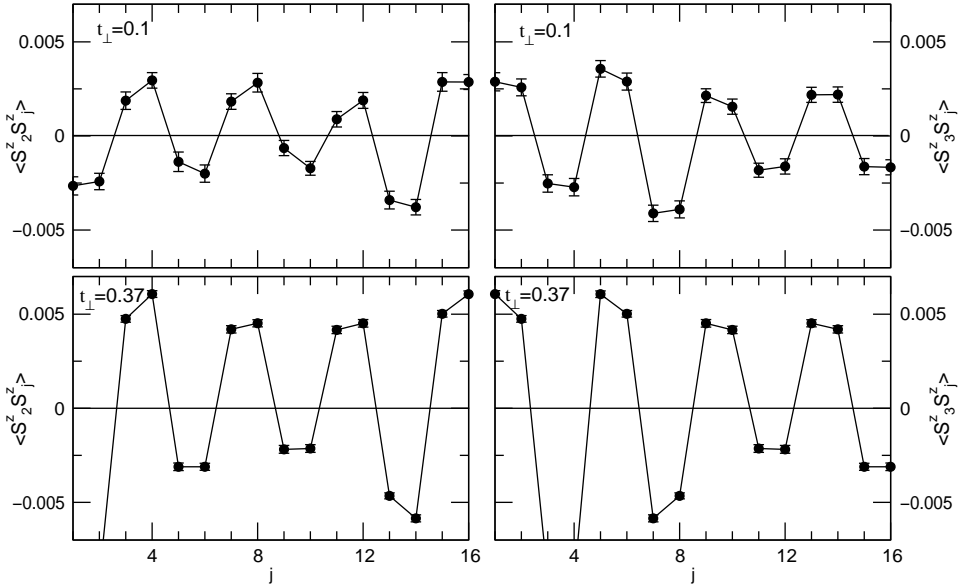}}
\caption{The z-z spin correlations for the 16$\times$6
dimerized dimer lattice. Correlations are shown between site 2 and
site 3 on the first chain and sites 1-16 on the second
chain with $U=6$, $V=1$ for two values of $t_{\perp}$ (0.1 and 0.37).
Lines are guides to the eye. }
\label{16x6_sdw_dd}]
\end{figure}
Physically, the reason for this persistence is the 
{\it coexisting} site CDW, whose nature is independent of $r_{4k_F}$ 
\cite{mrcc,umt}. 
We show in Fig.~\ref{delta_e_2kf4kf} the calculated $\Delta$E/$\Delta$E$_0$ 
for $r_{4k_F}$ = $r_{2k_F}$ (equal admixtures of 2k$_F$ and
4k$_F$ bond distortions), for the 8 $\times$ 2 
and 8 $\times$ 6 lattices for $U = 6$ and $V = 1$. 
The hopping integrals
corresponding to the distorted lattice here are 1.089, 0.974, 1.089
and 0.848, and the energy gained is being measured against the uniform
lattice. Starting from $t_{\perp}$ = 0.5, the one-electron $\Delta$E is
highly discontinuous. This is because distortions with $r_{4k_F} \neq 0$
do not correspond to a natural periodicity for the noninteracting system.
As a consequence the noninteracting wavefunctions are not suitable trial
wavefunctions for the CPMC calculation.
For the same reason the 16$\times$6 calculation
could not be performed here.
The similarities between the results for the 8 $\times$ 2 and
the 8 $\times$ 6 lattices are obvious. 
The ratio $\Delta$E/$\Delta$E$_0$
is independent of $t_{\perp}$ over a broad range of
$t_{\perp}$ and increases slightly for large $t_{\perp}$, 
indicating 
once again a stable 2D BCDW. Although only limited data could be obtained
for this case, the dimerized dimer lattice is very similar in character
to $r_{4k_F} \neq$ 0 (see Fig.~\ref{cartoon}(b)). In the following we show
convincing evidence
for persistent double-dimerization in 2D.

\subsubsection{The dimerized dimer lattice}

We have previously noted that Fig.~\ref{cartoon}(b) suggests that an 
alternate way to
view the BCDW/BCSDW states is as a dimer lattice with
additional structure within each of the dimer cells;
the dotted box in Fig.~\ref{cartoon}(b) represents one dimer.
Each dimer has one electron, leading to an ``effective
half-filled'' dimer band \cite{Bourbonnais,Kanoda,Kino,McKenzie,Gorkov}. 
Bond dimerization in the 1D 1/2-filled band is
unconditional for all $U > 2V$ \cite{review1,Dixit}, and thus this
dimer lattice itself distorts in a period 2 dimerization
pattern in 1D. In this section we show the additional 
result that the (anisotropic) 2D dimer
lattice is unconditionally unstable to a second
dimerization for all $t_{\perp}$.

We choose the hopping integrals between the two sites within
the dimer cell to be 1.2 in our calculations.
The two inter-dimer hopping integrals
for the uniform dimer 
lattice were taken to be 0.8, while for the
distorted ("dimerized") dimer lattice these were taken to be 0.7 and 0.9,
respectively ({\it i.e.}, the
dimerized dimer lattice has hopping integrals 1.2, 0.7, 1.2, 0.9
along each chain). Exact diagonalizations show that a $\pi$-phase shift 
between the chains ({\it i.e.}, dimer cells lying directly above each other, but
a strong inter-dimer bond on one chain facing a weak inter-dimer bond on the
next chain) gives the lowest total energy. Again we define $\Delta$E and 
$\Delta$E$_0$ as the electronic energies gained per
site upon interdimer bond 
distortion by the 2D and 1D lattices. 
Fig.~\ref{delta_e_dd} shows the $\Delta$E/$\Delta$E$_0$ 
behavior for the 8$\times$2 lattice over the complete range of $t_{\perp}$
and for the 8$\times$6 and 16$\times$6 lattices for several different
$t_{\perp}$ for
$U=6$ and $V=1$. The 8$\times$6 and 16$\times$6 lattices, taken together,
cover nearly the full
range of $t_{\perp}$, and the $\Delta$E/$\Delta$E$_0$ behavior
for these lattices closely follow the curve for the 8$\times$2 lattice.
As before, $\Delta$E/$\Delta$E$_0$ 
is significantly
greater than 1 for the complete range  $0<t_{\perp}<1$, indicating the
persistence of the dimerization of the dimer lattice in the interacting case,
whereas for the non-interacting case, the dimerization vanishes, as expected.

Fig.~\ref{16x6_sdw_dd} shows the interchain spin-spin correlations between
sites 2 and 3 on one chain and sites j = 1 -- 16 on a neighboring chain,
for a 16$\times$6 dimerized dimer system. 
Notice the far smaller contribution by the ...1010... intrachain charge
ordering here. This is because of the large difference between the hopping
integrals even in the ``uniform'' lattice with interdimer hopping integrals
of 0.8 here. Such a large bond dimerization diminishes the intrachain
...1010... contribution.
The spin-spin correlation
amplitudes cannot be directly compared to Fig.~\ref{16x6_spin_spin}
because of the different distortion amplitudes, but Fig.~\ref{16x6_sdw_dd}
shows that the SDW amplitude is significantly greater in the 
intermediate $t_\perp$ regime ($t_{\perp}$ = 0.37 in the Figure)
compared to the small $t_\perp$ regime
unlike the results in Fig.~\ref{16x6_spin_spin}.
Our calculations indicate that the larger the difference between the
intra-dimer and the
inter-dimer hopping integrals, the greater the range of the $t_{\perp}$
over which the SDW is stable. Thus with hopping integrals of 1.2, 0.9,
1.2 and 0.7 along each chain, the SDW in the 8$\times$6 lattice
persists even at $t_{\perp} = 0.6$ (in contrast to the 2k$_F$ bond-distorted
lattice of Fig.~\ref{cartoon}), 
but vanishes at still larger $t_{\perp}$. This is expected
from our discussion of the behavior of the SDW in Section III.D. Recall that
the smaller spin densities on the sites labeled `0' are influenced by both
the intrachain nearest neighbor as well as the interchain nearest neighbor
with opposite spin, and this competition creates a disordering effect. 
The larger the hopping integral between the `0' and the nearest intrachain
`1',
the larger the
$t_{\perp}$ necessary to create the disordering of the spin, hence
the greater stability of the SDW. We shall later argue that 
this same phenomenon is related to the very large magnetic moments
of the $\kappa$-(BEDT-TTF) salts. 

\subsubsection{Effects of additional Coulomb interactions} 

Fig.~\ref{cartoon} clearly suggests that interchain nearest 
neighbor Coulomb interaction $V_{\perp}$ stabilizes the BCDW 
further. We have confirmed this 
by exact numerical calculations for the 8 $\times$ 2 lattice, as shown in
Fig.~\ref{delta_e_vperp} below, where we have plotted 
$\Delta$E/$\Delta$E$_0$ for three different
values of $V_{\perp}$: 0, 0.5 and 1. Nonzero $V_{\perp}$ increases 
$\Delta$E further. Similar calculations were done also with variable
$V_{\perp}$ but fixed $V_{\perp}/t_{\perp}$. 
An even larger increase in $\Delta$E is found in this case. Implementing
$V_{\perp}$ over and above $V$ is difficult within the CPMC,
and therefore these calculations could not be performed for
larger lattices. However, based on 
\begin{figure}[htb]
\centerline{\epsfig{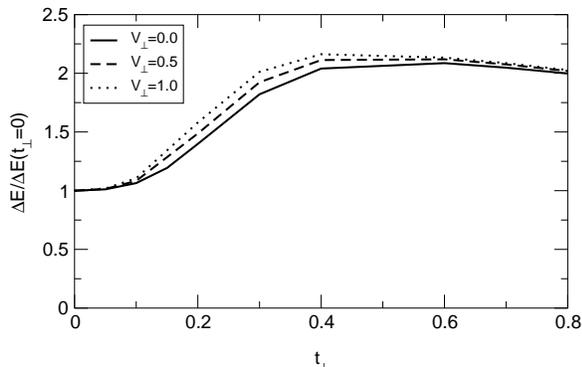}}
\caption{$\Delta$E/$\Delta$E$_0$ vs. $t_{\perp}$ for the 8 $\times$ 2
lattice, with $U = 6$, $V = 1$ and $V_{\perp} = 0, 0.5, 1.0$, for
$r_{4k_F}$ = 0.}
\label{delta_e_vperp}
\end{figure}
the similarities
between the $\Delta$E behavior
of the three lattices studied in Figs.~\ref{delta_e_2kf} and 
\ref{delta_e_dd}, no difference in the larger
lattices is expected.

\section{Summary of theoretical results}

We have performed detailed numerical calculations of various broken symmetries
for the 2D 1/4-filled band within Eq.~(1) and for the effective 1/2-filled 
band of dimer lattice within Eq.~(2), for $U$ = 6,
$V$ = 1. Regarding these parameter values, the broken symmetries
we have found will occur for all intermediate to strong $U$ but
require $V$ to be less than a critical $V_c \geq 2t$ \cite{HF}.

We have discovered three distinct new results in 1D.
First, we have confirmed that the BCDW state occurs spontaneously even
for zero e-ph couplings (see Fig.~\ref{1d-fig}(a)). 
The bond distortion pattern in the
center of a long open chain corresponds to a pure 2k$_F$ distortion, and
coexists with the 2k$_F$ ...1100... type charge modulation. 
Second, we have shown that a BOW appears spontaneously in a uniform periodic
ring when the  
SDW $\uparrow,\downarrow$,{\tiny{$\downarrow,\uparrow$}} is superimposed,
confirming
the synergetic cooperation between e-e and e-ph interactions. 
The BOW pattern corresponds to r$_{4k_F}$ = 0 (see Eq.~(7))
when the amplitude of the
superimposed SDW is relatively weak (Fig.~\ref{1d-fig}(b)), 
but switches over to r$_{4k_F} \neq 0$ when the SDW amplitude is large
(Fig.~\ref{1d-fig}(c)). Our earlier demonstrations of the
BOW-SDW coexistence were only for the bond
distorted periodic systems.
Finally, from exact calculations for a periodic dimerized
dimer ring, we have established the new result
that the BOW here also coexists with
the ...1100... 2k$_F$ CDW, with the large (small)  
charges occupying the sites connected by the stronger (weaker)
interdimer W$'$ (W) bond (see Fig.~\ref{1d-fig}(d)). 
Our earlier work had claimed that a 1/4-filled
description was essential to obtain the BCDW and the BCSDW states. As shown in
Fig.~\ref{1d-fig}(d), the same result is obtained, however, even for the
dimer lattice, {\it provided} the second dimerization is allowed to occur.

Three different bond distortion patterns were investigated in 2D. These
correspond to r$_{4k_F}$ = 0 (Fig.~\ref{cartoon}(a)), 
r$_{4k_F}$ = r$_{2k_F}$ (Fig.~\ref{cartoon}(b)), 
and the dimerized dimer lattice. In all cases a $\pi$-phase shift in the bond
distortion between consecutive chains gives the lowest energy. From 
calculations of energy gained upon bond distortion, we conclude that 2D 
bond distorted lattices with r$_{4k_F}$ = 0 and r$_{4k_F}$ = r$_{2k_F}$ are
both more stable than the uniform lattice (see numerical results in
Figs.~\ref{delta_e_2kf} and \ref{delta_e_2kf4kf}). 
Similarly, the dimerization of 
the dimer lattice is also unconditional (see numerical results in 
Fig.~\ref{delta_e_dd}). 
The persistence of the distortions is a novel effect of e-e interactions and
is in contradiction to what is expected within one-electron nesting concepts.
The ground state of the strongly correlated 1/4-filled band is 
therefore a novel insulating BCDW state for all $t_{\perp}$. 

The persistence of the BCDW for all
anisotropies is also evident from the charge density calculations.
In Fig.~\ref{delta_rho}, we have shown the amplitude of the 
CDW that accompanies the 
r$_{4k_F}$ = 0 BOW as a function of $t_{\perp}$. In the absence of e-e
interaction, the CDW amplitude decreases rapidly with $t_{\perp}$ even with
nonuniform hopping integrals.
One interesting aspect of these calculations
is that the CDW pattern is the same for all bond distortion patterns.
Our computer capabilities do not allow us to 
determine self-consistently which of the three BOW patterns dominate within
Eqs.(1) and (2) for a given $U$, $V$, $t_{\perp}$, $\alpha$ and $\beta$. 
This is, however, largely irrelevant, because the charge ordering is the
same with all the bond distortion patterns.

The SDW behavior is different from those of the BOW and the CDW. As seen from
our numerical calculations of interchain spin-spin correlations
in Figs.~\ref{8x6_spin_spin} and \ref{16x6_spin_spin}, the SDW amplitude 
of the novel BCSDW state is initially 
enhanced by $t_{\perp}$, but with further increase in $t_{\perp}$ the 
SDW vanishes, indicating a singlet BCDW state again
in the large $t_{\perp}$ region.
The range of $t_{\perp}$ within which a stable SDW is found depends 
on the BOW pattern, and within the dimerized dimer lattice 
(see Fig.~\ref{16x6_sdw_dd}) the
SDW can be stable over a wider range of $t_{\perp}$.

\section{Comparison to Experiments on the Insulating States in 2:1 organic CTS}

Experimentally, the organic cationic CTS, with 
cation:anion ratio of 2:1, span the range
$t_{\perp} \leq 0.1$ in (TMTTF)$_2$X to $t_{\perp}$ $\sim$ $1$ in certain
(BEDT-TTF)$_2$X. Hence these materials provide a critical testing
ground for our theoretical results. In reference [24], we 
compared our theoretical predictions regarding the BCSDW state to
the mixed CDW-SDW found experimentally in (TMTTF)$_2$Br, (TMTSF)$_2$PF$_6$
and $\alpha$-(BEDT-TTF)$_2$KHg(SCN)$_4$. Here we make additional, 
more detailed 
comparisons, distinguishing between 1D TMTTF and weakly 2D
TMTSF-based compounds, and
also emphasizing the similarities and differences between the salts of 
BEDT-TTF and BETS 
with different crystal structures.
In the case of the TMTTF and TMTSF band structure calculations of hopping 
integrals have been been summarized by Yamaji \cite{Yamaji1}. In both cases
the lattice is anisotropic triangular in nature, which would correspond
to our rectangular lattice with one additional diagonal hop $t_{diag}$ 
beyond the usual $t_{\perp}$. Both $t_{\perp}$ and $t_{diag}$ are small in
the 1D TMTTF, while they are comparable in TMTSF and about 0.1$|t|$ in 
magnitude.  
As discussed in section III.D, the paired electron crystal
ordering even along the diagonal directions in the configurations shown in 
Figs.~2(a) and (b) indicate that the BCDW and the BCSDW states continue to be
stable for nonzero $t_{diag}$ and there is thus no loss of generality in
considering a rectangular lattice. 
Several crystal structures occur in the
BEDT-TTF systems, and more subtle and individual analyses for the 
different cases are required.
Our aim is to show that a variety of recent experiments indicate
that the BCSDW and the BCDW are appropriate descriptions of the
insulating states of this entire class of 2:1 cationic CTS, and conversely,
the very nature of the insulating ground state in certain cases provides
direct verification for some of our more surprising theoretical results.
We discuss below each class of material individually.

\subsection{(TMTTF)$_2$X}

The (TMTTF)$_2$X compounds are nearly 1D
semiconducting materials with weak to moderate dimerization along the
stacks at high temperature. Because of this dimerization, they have often been
described within the effective 1/2-filled band picture 
\cite{Bourbonnais,Bourbonnais2}.
Further dimerization of the dimerization occurs
below the SP transition temperature T$_{SP}$ ($\sim$ 15 K).
Existing theories of the SP transition in these systems
\cite{Bourbonnais2} do not discuss the simultaneous appearance of the 2k$_F$
CDW and {\it assume} that the site populations 
continue to be uniform below T$_{SP}$. 
As depicted in Fig.~\ref{eff_half}(c), and as confirmed in
Fig.~\ref{1d-fig}(d), 
independent of whether these systems are considered
as 1/4-filled or effective 1/2-filled with a dimer lattice, the appearance
of this 2k$_F$ CDW is unconditional and the site populations are
therefore not uniform. In a recent NMR study of 
$^{13}$C spin-labeled (TMTTF)$_2$PF$_6$ and (TMTTF)$_2$AsF$_6$
charge-ordered states have been found \cite{Brown}. 
Although such a charge-ordering suggests agreement with the theory presented
here, one problem is that the initial appearance of the charge-ordered phase
(at $\sim$ 70 K in (TMTTF)$_2$PF$_6$)
occurs considerably above T$_{SP}$ (15 K) \cite{Brown}.
There are two possible reasons why the charge-ordering might appear at
a temperature T$_{CO} >$ T$_{SP}$. First, this might be due to fluctuation
effects associated with the 1D nature of the crystals.
As has been shown by Schulz \cite{Schulz1}, fluctuation effects associated
with the SP transition may be seen at temperatures as high as 4T$_{SP}$, in
which case signatures of charge ordering would also become visible at these
high temperatures. The observation of diffuse X-ray scattering at 2k$_F$
in this material already at $\sim$ 60 K \cite{Pouget1,Pouget2} 
seems to support this possibility.
A second possibility is that the charge-ordering is driven primarily by the
Holstein e-ph coupling $\beta$ in Hamiltonian (1), and the SSH coupling
$\alpha$ is small, such that actual lattice displacement and spin singlet
formation takes place at lower temperature. Independent of which mechanism
dominates to give T$_{CO} >$ T$_{SP}$, it is important to keep in mind that
(a) no charge-ordering is expected at all within conventional theories of
SP transition, and (b) as discussed extensively in section III,
charge ordering of the type ...1010..., as has sometimes been suggested (see
below and footnote \onlinecite{HF}), promotes equal intrachain bonds, and 
therefore {\it the SP
transition could not occur if the ...1010...charge-ordering had 
taken place.}
Finally as has been pointed out by us previously \cite{umt},
charge-ordering of the 
type ...1100... also occurs in the
SP phase of the anionic 1:2 TCNQ solids. 

Although most (TMTTF)$_2$X exhibit the SP transition, the material 
(TMTTF)$_2$Br exhibits a transition to a SDW \cite{Coulon,Torrance}, 
like the (TMTSF)$_2$X. Also like
the (TMTSF)$_2$X, this material can become superconducting, although at a
relatively high pressure of 26 kbar. Within the structural
classification scheme described by
Jerome \cite{Jerome}, this difference is due to the larger $t_{\perp}$
in (TMTTF)$_2$Br (relative to the other TMTTF).
We therefore discuss this material
along with the (TMTSF)$_2$X.

\subsection{(TMTTF)$_2$Br and (TMTSF)$_2$X}

X-ray scattering studies by
Ravy and Pouget \cite{Pouget1,Pouget2} have shown that in both (TMTTF)$_2$Br
and the prototype TMTSF system, (TMTSF)$_2$PF$_6$, CDW distortions occur
below the SDW transition temperature T$_{SDW}$. Similar conclusions have
been reached also by Kagoshima et al. \cite{Kagoshima}. In (TMTTF)$_2$Br
evidence for a 4k$_F$ lattice instability was found \cite{Pouget1,Pouget2}, 
clearly suggesting that
the insulating state here is the BCSDW of Fig.~\ref{cartoon}(b). 
In (TMTSF)$_2$PF$_6$
the authors claim a ``purely electronic CDW'', which would indicate the
dominance of the 2k$_F$ CDW over the BOW. 
Since, however, in both the 1/4-filled band
and the effective 1/2-filled band, the 2k$_F$ CDW necessarily coexists with
a BOW, the experimental work merely indicates that the transition to the
BCSDW state is driven mainly by the Holstein e-ph
coupling in Eq.~(1) rather than the SSH coupling ({\it i.e.}, $\alpha$ is 
small), so that the actual modulations of the intermolecular distances
are small \cite{KOY}. This would agree with one of the two possible reasonings
given by us for T$_{CO}$ being larger than T$_{SP}$ in (TMTTF)$_2$PF$_6$
and (TMTTF)$_2$AsF$_6$, as discussed above.

\begin{figure}[htb]
\centerline{\epsfig{width=3.0in,file=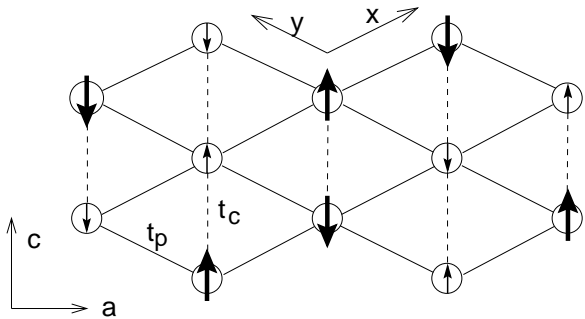}}
\caption{Schematic view of structure of $\alpha$-(BEDT-TTF) donor plane
from Mori et al. \cite{Mori} and Ducasse and Fritsch \cite{Ducasse}.
The solid lines correspond to stronger interstack $t_p$ hopping integrals,
the dotted lines to weaker intrastack $t_c$ hopping integrals.
The $a$ and $c$ directions indicated are the crystal axes,
and the $x$ and $y$ directions correspond to along the chain
and perpendicular to the chains in Fig.~\ref{cartoon}.
The arrangement of the spins in the BCSDW state
is indicated. Any SDW should be weak because of the nearly isotropic 2D
nature of the lattice, but nonvanishing because of the nonzero
$t_c$, which becomes $t_{diag}$ in the $x$-$y$ coordinate system of
Fig.~2 (see text).}
\label{fukuyama_fig}
\end{figure}
One additional comment appears to be necessary. Fr\"ohlich mode sliding
conductivity has been seen in (TMTSF)$_2$X \cite{Gruner2}. 
While this indicates
a weak incommensurability of the density wave (see below), an equally
important point is that the sliding conductivity in the past has been
ascribed to a SDW: the SDW collective transport is viewed as
that of two CDWs, one for each spin subband. The actual displacement of the
charge density
 is difficult to visualize in configuration space within this  
picture. We believe that the experimental demonstration of the
coexisting CDW and the present theoretical work, taken together, suggest the 
more coherent viewpoint that the sliding mode conductivity is that of
a BCSDW.

\subsection{$\alpha$-(BEDT-TTF)$_2$MHg(SCN)$_4$}

 This class of materials,
with M = K, Rb, Tl and NH$_4$ has been of considerable interest
recently. M = NH$_4$ is a superconductor, but M = K, Rb, Tl are
non-superconducting. Early magnetic susceptibility 
studies in the M = K
material had indicated anisotropic susceptibility below the so-called
``kink'' transition that occurs at 10 K, indicating a SDW; here the kink refers to the change
in slope that occurs in the temperature dependence of the resistivity and
the Hall coefficient. On the other hand, 
analysis of the angle-dependent magnetoresistance oscillations by
Sasaki and Toyota led these authors to conclude already in 1995, prior to the
experiments by Pouget and Ravy in the (TMTSF)$_2$PF$_6$, that
the dominant broken symmetry in $\alpha$-(BEDT-TTF)$_2$MHg(SCN)$_4$ is a 
CDW \cite{Toyota}. Since, however, a CDW would not explain the anisotropic
susceptibility, Sasaki and Toyota concluded that the broken symmetry here
is a ``mysterious'' state that is a ``SDW accompanied by a CDW'' or a 
``CDW accompanied by a SDW''. Muon spin resonance 
studies indicate very small magnetic
moment per BEDT-TTF molecule here, $\sim$ 0.003 $\mu_B$ \cite{Pratt}
(to be compared against
0.08 $\mu_B$ in (TMTSF)$_2$X \cite{Wzietek} and 0.4 -- 1 $\mu_B$ per
BEDT-TTF dimer in $\kappa$-(BEDT-TTF)$_2$Cu(CN)$_2$Cl \cite{Kanoda1}, see 
below). More recent $^{13}$C-NMR studies in the M = Rb indicate even
smaller magnetic moment (if it exists at all) $\sim$ 1 $\times$ 10$^{-4}$ $\mu_B$ 
\cite{Miyagawa}. Recent theoretical \cite{Harrison} 
and experimental \cite{Biskup} investigations conclude either that 
the dominant
broken symmetry here is a CDW or that it is not a
conventional SDW \cite{Kuhns}.

We point out here that a mixed state with very small magnetic moments is 
exactly what is expected within our theory. In Fig.~\ref{fukuyama_fig}
we have given a
schematic view of the structure of the donor plane in
$\alpha$-(BEDT-TTF)$_2$MHg(SCN)$_4$. The one-electron hopping integrals
(called ``t$_{p}$'' and ``t$_{c}$'' in the figure) have been calculated using 
approximate one-electron techniques by
Mori et al. \cite{Mori} and Ducasse and Fritsch \cite{Ducasse}. 
Here the t$_{p}$ correspond to the interstack hopping and the t$_{c}$ to the
intrastack hopping.
Four slightly different
p-type integrals and three slightly different c-type integrals are obtained by these
authors. We ignore the small differences within each type of hopping 
integrals, as a more important effect is the periodic modulation that appears
with the BCDW.
We believe that what is relevant in the
present context is that $t_{p} > t_{c}$.
The $\alpha$-BEDT-TTF lattice is then simply a rotated (by approximately 
45$^o$) version of our rectangular lattice with both $t$ and $t_{\perp}$ = 
$t_p$ and $t_{diag} = t_c$.
Our calculations 
(see Figs.~\ref{delta_e_2kf}, ~\ref{delta_rho}, ~\ref{delta_e_2kf4kf} and
~\ref{delta_e_dd}) 
show that even
at $t_{\perp} \sim 1$ the correlated 1/4-filled band (or the dimerized
dimer lattice) remains bond and charge-distorted, 
while based on the ..1100... ordering
along the diagonals we have argued that $t_{diag}$ does not destroy this
order (see section III.D). 
Furthermore, while $t_{\perp} > t_{\perp}^c$ destroys the SDW order 
(leaving the BCDW intact) by disordering the spins on the sites labeled
`0' (see section III), a small $t_{diag}$ will have a tendency to restore
it, since now each small spin has two neighbors with spins of the same sign
and one spin with opposite sign. Thus, the experimentally observed strong BCDW and a weak nearly
vanishing SDW is exactly what we expect within our theory.
Further evidence for
a partial gap has been found in the $^{13}$C-NMR studies of 
$\alpha$-(BEDT-TTF)$_2$KHg(SCN)$_4$ in high magnetic fields, in a region where
the system was previously thought to be a metal\cite{Kuhns}. 
In Fig.~\ref{fukuyama_fig} 
we give a schematic of the spin
arrangement in the $\alpha$-BEDT-TTF lattice; note that the underlying 
$x \leftrightarrow y$ symmetry in
the isotropic 2D limit
implies that there are two
degenerate orthogonal 2D BCDW states here.

Since in $\alpha$-(BEDT-TTF)$_2$MHg(SCN)$_4$ charge-ordering has also
been discussed by Kino and Fukuyama \cite{Kino}, and more recently, by Seo 
\cite{Seo}, we should point out
that the charge-ordering proposed by these authors is different from that
in Fig.~\ref{fukuyama_fig}. Our charge-ordering in Fig.~\ref{fukuyama_fig}
is a rotated version of Fig.~\ref{cartoon}, where the occupancy scheme
is ...1100... along the x-direction and along the diagonals. The 
charge-ordering found by Kino and Fukuyama, and by Seo, assumes that the
...1010... order dominates over the ...1100... order. The ordering 
determined by Kino and Fukuyama is within a Hartree-Fock solution to the
simple Hubbard model (zero intersite Coulomb interaction and zero
e-ph coupling) and consists of a stripe structure with 
stack occupancies (c-direction in Fig.~\ref{fukuyama_fig}) alternating,
{\it i.e.}, stacks are either completely filled or 
completely devoid of holes).  
More recently, Seo has repeated these calculations by incorporating nearest
neighbor Coulomb interaction $V$, but by treating $U$ within the Hartree-Fock
approximation and the $V$ within the Hartree approximation. Different
stripe structures, including that of Fukuyama and Kino, are found now, but
once again, these are derived fundamentally from the occupancy scheme
...1010... As has, however, been pointed out by previous authors 
\cite{Hirsch2,umt}, the ...1010... charge ordering for the case
of $V$ = 0 is an artifact of the
Hartree-Fock approximation. Similarly, the Hartree approximation for $V$
also exaggerates the ...1010... order while the Hartree-Fock treatment of the
Hubbard term exaggerates the SDW order \cite{HF}.
This is precisely why these authors
find very large magnetic moments in the $\alpha$-phase materials, in 
disagreement with experiments.

\subsection{$\kappa$-(BEDT-TTF)$_2$X}

The deviation from the rectangular
lattice is much stronger here \cite{Ishiguro}. Crystal structure effects
are very strong, and as a consequence the lattice is strongly dimerized, with
the dimer sites forming an effective triangular lattice \cite{Kino}. 
The strong deviation
from the rectangular lattice precludes direct comparisons against our theory.
A more elaborate discussion of the spin arrangement will be given elsewhere.
Here we only point out that (a) our calculations with the dimerized dimer
lattice indicate that very large spin moments are possible when the
intra-dimer hopping integrals are large compared to the inter-dimer hopping
(see Fig.~\ref{16x6_sdw_dd}),
in qualitative agreement with the observed very large
magnetic moment in $\kappa$-(BEDT-TTF)$_2$Cu(CN)$_2$Cl \cite{Kanoda1}, and
(b) each dimer of BEDT-TTF molecules has the cartoon occupancy of 10 or 01
and the ...1100... ordering along one direction and
...1010... ordering along another (see Fig.~\ref{cartoon}), thereby 
reducing the spin frustration among the dimer sites 
forming the triangular lattice. In the absence of this population difference
within each dimer cell (and the population difference is a consequence only
of dimerization of the dimer lattice) 
the frustration within the triangular lattice would have 
severely reduced magnetic moments.
We further point out that a
pseudo-gap in the spectrum of magnetic excitations
has been observed in the SDW phase of 
$\kappa$-(BEDT-TTF)$_2$Cu[N(CN)$_2$]Br 
\cite{Mayaffre,Kawamoto,Nakazawa}
and
$\kappa$-(BEDT-TTF)$_2$Cu(NCS)$_2$ \cite{Kawamoto};
this is in agreement with the dimerization of the
dimer lattice, since without the 
second dimerization there should be no spin gap
within the 2D antiferromagnet.

The material $\kappa$-(BEDT-TTF)$_2$Cu$_2$(CN)$_3$
merits separate discussion. This material is not antiferromagnetic, and
measurement of spin susceptibility due to the BEDT-TTF components exhibits
a steep drop below 10 K, suggesting SP-like behavior \cite{Komatsu}. This
behavior is very similar to that in the BETS-based materials, which we discuss
below, where we point out that for $\rho$ = 1/2, this behavior is expected
for the case of large $t_{\perp}$ ($> t_{\perp}^c$).

\subsection{$\lambda$-(BETS)$_2$GaBr$_z$Cl$_{4-z}$ (BETS = BEDT-TSF)}
 
These materials, discovered
only recently \cite{Kobayashi,Tanaka,Montgomery}, are 
superconducting for
$0 < z \leq 0.8$ and semiconducting for $0.8 < z < 2.0$.
Thus the proximity
between a semiconducting and a superconducting state that characterizes the
TMTSF and the BEDT-TTF is also a characteristic feature of the 
$\lambda$-BETS. In contrast to the TMTSF and the BEDT-TTF systems, however,
the semiconducting state in the BETS is {\it nonmagnetic} and possesses a
spin gap \cite{Kobayashi2}. Magnetic susceptibility studies indicate absence
of anisotropy in the susceptibility, and no spin-flop transition (signature
of antiferromagnetism) was found down to 10 K, which is close to the 
maximum superconducting critical temperature T$_c$
(onset 7.5 K, and even higher
in certain samples) \cite{Kobayashi}. The absence
of the SDW is particularly perplexing here in view of the strong
two-dimensionality predicted within extended H\"uckel band calculations
\cite{Montgomery}.

The lattice structures of the $\lambda$-(BETS)$_2$GaBr$_z$Cl$_{4-z}$ are
known \cite{Kobayashi}. The stacking of the organic donor molecules is 
very similar to the
$\beta$-BEDT-TTF systems, {\it i.e.}, a nearly rectangular lattice with strong
intrastack coupling, weaker transverse coupling, and very weak coupling
along one diagonal. The nearly rectangular lattice permits comparison
with our theory. One interesting feature of the lattice structure is that the
intrastack bonds have strengths that are W$'$SWS, 
exactly the structure expected for the $r_{4k_F} \neq 0$ lattice in
Fig.~\ref{cartoon}(b) as well as the dimerized dimer lattice. We believe
that while the difference between the strong and weak bonds is a crystal
structure effect, the further dimerization of the dimer lattice is a 
consequence of the BCDW instability discussed here.

Hartree-Fock
calculations by Seo and Fukuyama \cite{Seo2} 
within an anisotropic Hubbard Hamiltonian
gave an antiferromagnetic ground state 
instead of the nonmagnetic state. Since Hartree-Fock calculations overestimate
antiferromagnetism, these authors then chose the
$U \to \infty$ limit of Hubbard model to arrive at a  
dimerized, anisotropic 2D Heisenberg
spin Hamiltonian, each lattice site of which corresponds to one dimer of
the original BETS lattice. The 
antiferromagnetic-SP boundary within the 2D dimerized Heisenberg spin
Hamiltonian has been investigated 
by Katoh and Imada using QMC simulations \cite{Imada}. For the longitudinal
and transverse exchange integrals derived by Seo and Fukuyama, the QMC
calculations still predict the antiferromagnetic structure \cite{Seo2}.
Seo and Fukuyama
explain the spin gap 
in $\lambda$-BETS by claiming that the second dimerization of the dimer 
lattice ({\it i.e.}, intermolecular distances W$'$SWS, 
instead of WSWS) takes these
systems to the 
1D side of the 1D-2D antiferromagnetic-SP boundary, exactly as
(TMTTF)$_2$PF$_6$, even though the actual transverse hopping integrals are
large.

We believe that the problem faced by these authors arises entirely from
their effective 1/2-filled band approximation. As seen in 
Fig.~\ref{16x6_sdw_dd}, the dimerization of the dimer lattice enhances the
SDW in the region of small to intermediate $t_{\perp}$ and therefore
cannot be the origin of the spin gap or supposedly 1D behavior. 
Recall also that
(TMTTF)$_2$PF$_6$, which is certainly on the 1D
side of the 1D-2D boundary, is nonsuperconducting.
In contrast,
$\lambda$-BETS does become superconducting and that too at a $T_c$ that
is considerably higher than that in the
(TMTSF)$_2$X, indicating what we believe to be strongly 2D character
\cite{Montgomery}.
We believe that the solution to this puzzle lies in recognizing the $\rho$
=1/2 character of the (BETS)$_2$X.
An essential difference between the effective 1/2-filled band model of
Seo and Fukuyama and ours is that within the former, there are only
two regions, nearly 1D and 2D, with the spin states as singlet and
antiferromagnetic, respectively. Our work indicates that there are three
distinct regions, singlet, antiferromagnet, and singlet again, as a function of
increasing $t_{\perp}$, independent of whether one assumes a 1/4-filled band or an
effective 1/2-filled band.
We therefore believe that a more natural explanation of the spin gap phase
is obtained within our theory, with 
the singlet
ground state in semiconducting BETS not being due to $t_{\perp}$ that is too
small, {\it but due to a $t_{\perp}$ that is too large ($> t_c$)}
to give SDW. This would
be in agreement with the strong two-dimensionality of these systems 
\cite{Kobayashi,Montgomery}. We believe that the same explanation also
applies to the $\kappa$-(BEDT-TTF)$_2$Cu$_2$(CN)$_3$, discussed in the
above. We predict that experiments that can probe
charge ordering will find two kinds of BETS molecules with different
electronic populations, with greater charge densities on the two BETS molecules
that are linked by the W$'$ bond.

\section{Possible Implications for Organic Superconductivity}

What might be the implications of our BCSDW and BCDW states to
organic superconductivity, the mechanism for which
remains unclear despite two decades of research? We present
here several partial responses to this challenging
question.

First, given the
the robustness of the BCDW/BCSDW in the exactly 1/4-filled band,
we believe that the superconductivity must be the result of
weak incommensurability in the actual materials. Specifically,
we suggest and discuss in more detail below, that superconductivity
arises from the pairing of commensurability defects in the
background BCDW/BCSDW. That such weak
incommensurability exists
is strongly indicated by (i) the observation of a zero-energy mode
in the optical conductivity \cite{schwartz,vescoli}
of (TMTSF)$_2$PF$_6$ and (TMTSF)$_2$ClO$_4$;
(ii) the observation of Fr\"ohlich mode sliding transport in the 
same materials \cite{Gruner2};
and (iii) the observation of a ``partially gapped Fermi surface'' 
in the metallic
region \cite{Kuhns} of $\alpha$-(BEDT-TTF)$_2$KHg(SCN)$_4$. 
Extremely interesting results in this context were reported by Komatsu
et. al. \cite{Komatsu}, who showed that the superconductivity in 
$\kappa$-(BEDT-TTF)$_2$Cu$_2$(CN)$_3$ was due to a subtle change in the
valence state of the Cu. The pure $\kappa$-phase material is a semiconductor
with the Cu valence of +1. According to the authors of 
Ref. \onlinecite{Komatsu}, the superconducting
phase corresponds to a different material ($\kappa'$ in the authors' notation)
in which some of the Cu (several hundred ppm) have acquired valency 2+. 
This was confirmed from ESR studies. The
increase in Cu valency decreases the overall negative charge on the anion,
and therefore the overall positive charge on the cation, providing
a weak incommensurability that appears to be 
essential for superconductivity \cite{Komatsu}.
This result lends credence to our suggestion
that organic superconductivity arises from the
pairing of commensurability defects within the BCDW/BCSDW background.

Second, the similarities
between the organic and high temperature oxide superconductors have been
pointed out in recent years by several research groups 
\cite{McKenzie,SM,Greene,Uemura,Brandow}.
One obvious apparent similarity between these two classes
of superconductors is the proximity of the SDW to 
superconductivity. Our studies suggest that superconductivity
in the organics is
actually occurring at the interface of a Coulomb-induced BCDW that for a 
range of $t_{\perp}$ coexists with the SDW. 
It therefore seems more likely
that the pair binding is actually driven by the BCDW,
and not the SDW, although it is probable that the symmetry of the pairing
state may depend on the SDW (see below).
As noted above, the experimental
observation of superconductivity in
the $\lambda$-(BETS)$_2$GaCl$_4$ (where no proximate SDW is 
observed \cite{Kobayashi2}) supports this view.
An important implication of this perspective is that
it casts doubt on recent spin-fluctuation theories of {\it organic} 
superconductivity within the effective 1/2-filled correlated electron
model \cite{Moriya1,Kontani,Schmalian,Vojta,Aoki}. The consequences
of this conclusion from the organics for the high $T_c$ materials
are unclear, but it is perhaps
not irrelevant in this context to point out that evidence for 
superconductivity within the 2D nearly 1/2-filled Hubbard model, which
for large $U$ has strongly AFM behavior, has remained
elusive\cite{Gubernatis1,Gubernatis2,Lin}, despite more than a
decade of intense research\cite{Dagotto,Scalapino}.

Third, there are striking
similarities between this ``doped'' BCSDW/BCDW scenario and several
other theoretical suggestions of superconductivity
induced by doping of exotic ``paired'' semiconductors.
As we have noted previously, the BCSDW and the BCDW states
are very similar to the ``paired electron crystal'' (as opposed
to the monatomic Wigner crystal) found by Moulopoulos and 
Ashcroft for the intermediate density electron gas \cite{moulo}. 
Superconductivity near
the ``melting'' transition of the paired electron crystal has been
conjectured by a number of authors in the past
\cite{Chester,Andreev,Nelson,Leggett}, even before
the discovery of organic or high T$_c$ superconductivity.
The commensurate BCDW is also qualitatively similar
to a ``negative U - positive V'' effective 1/2-filled extended Hubbard model,
with the effective lattice sites sites consisting of 
(a) the ``occupied'' pair (`1--1') of nearest neighbor sites, 
and (b) the ``unoccupied'' pair (`0--0') of nearest neighbor sites, in 
Fig.~\ref{eff_half}(c). Within this scenario, there is an effective attraction
between the carriers on the ``occupied'' pair of dimer sites, but an effective
repulsion between two pairs of occupied dimers.
For models of this type, it is known
that diagonal and 
off-diagonal long-range order can in principle
coexist slightly away from commensurate filling \cite{Alexandrov,crf,Aubry99}.
Further, Imada has studied \cite{Imada1} a 2D spin-Peierls state 
(not possible in the
monatomic 1/2-filled band) in which each composite site is again a dimer, 
with the dimer sites now having occupancies
`10' and `01' (see Fig.~\ref{cartoon}(b) and note that the 
bonds between a `10' and `01' and between a `01' and a `10' are
different, giving rise to a spin-Peierls-like behavior).
His numerical simulations find evidence for superconductivity
in the hypothetical doped 2D spin-Peierls state \cite{Imada1}. 
Finally, Emery, Fradkin, and Kivelson have recently suggested \cite{efk00} 
that superconductivity
can exist for incommensurate fillings in models that support
stripe phases and in which a spin gap is present. Since the
analysis in Ref.~\onlinecite{efk00}
does not make direct contact with an initial
microscopic Hamiltonian, but rather
posits the form of the effective Hamiltonian in the vicinity of an
unpinned stripe phase, it is not possible immediately to make
detailed comparisons with our results. We can, however, make
two comments. First, Ref.~~\onlinecite{efk00} reflects the widespread
belief that models within which a spin gap persists in the
doped state are strong candidates for a microscopic theory of correlated 
superconductivity. Our preliminary numerical evidence suggests
that both the BCDW and the BCSDW will continue to have
a spin gap when doped; further work is in progress to confirm this.
Second, regarding the attractive possibility that our BCSDW/BCDW
state provides the background charge order within which
commensurability defects may pair to form a superconducting
state, we note that 
the occupancy schemes in Fig.~\ref{eff_half}(c) and 
Fig.~\ref{cartoon}(a) and (b) resemble intersecting stripes, where each stripe
is obtained by connecting the `1--1' bonds along the x- and x+y (--x+y) 
directions.

The possible BCSDW/BCDW to superconductor transition in the organic
CTS clearly requires further study. We close our present discussion of 
this topic
with comments on three important open issues: (i) the possible mechanism
for superconducting pairing; (ii) the problem of phase separation; and
(iii) the symmetry of the order parameter.

First, the possible mechanism for 
pairing of commensurability defects within the 2D BCDW
can be visualized most simply in the rigid bond limit,
where nearest-neighbor bonds retain their individual distortions 
independent of the
occupancies of the sites linked by these bonds. The commensurate BCDW in
this limit can be viewed as consisting
of ``quasimolecules'', where each quasimolecule is a 
``1-1'' dimer. If two holes are now removed from the system, it is
energetically preferable to destroy one ``quasichemical bond,'' thereby
creating an intersite (small) bipolaron, as opposed to destroying two bonds
and creating two polarons. Thus, within the W$'$SWS structure
 ($t_S > t_{W'} > t_W$), each W$'$ bond acts as a ``negative-U''
center in the rigid bond limit. It is of course highly unlikely that
superconductivity can be obtained, at least at the experimental T$_c$,
due to condensation of small bipolarons \cite{crf}, so this might
appear to present a serious problem for this proposed mechanism. In fact,
when one goes beyond the oversimplified rigid bond limit to the full
model that correctly reflects the cooperation between e-e and e-ph
interactions in the 1/4-filled band, one finds that the actual
commensurability defects are more like the extended, ``resonant''
(and therefore mobile) bipolarons that are indeed candidates
for explaining superconductivity in strongly 
correlated systems \cite{crf,Aubry99}.
To understand this in detail, consider again the weakly
incommensurate BCDW, starting from the 1D limit,
but now with the e-ph interactions included. Below the 4k$_F$ transition
temperature T$_{4k_F}$, but above the 2k$_F$ transition temperature
T$_{2k_F}$, 
incommensurability leads to {\it fractionally charged} solitons with
charge e/2, and each vacancy creates two such defects \cite{Rice,Kivelson1}.
Previous work has assumed that the soliton charge remains e/2 even below
the 2k$_F$ transition, which implies that two vacancies create four such
defects \cite{Kivelson1}. However, Ref. \onlinecite{Kivelson1}
assumes that the site charge
density remains uniform even below the 1D 2k$_F$ SP (dimerization of the dimer
lattice) transition, which is precisely what we have shown here not to
be the case. Indeed, as a consequence of this spatial charge
inhomogeneity (charge ordering), the ``solitons'' now acquire
integer charge ({\it i.e.}, two fractionally charged solitons bind to give a
single soliton with charge +e), as we have shown explicitly elsewhere
\cite{unpublished}. A pair of added vacancies within the 1D BCDW below
T$_{2k_F}$ therefore creates (only) two solitons. In the strictly 1D 
limit, these
do not bind, but with increasing $t_{\perp}$, one expects binding to a
{\it large} bipolaron. The source of this binding is precisely
the same as the source of soliton confinement in coupled chains
of polyacetylene \cite{review1}: in the region between the two
defect centers the phase relationships between the BCDW's on neighboring
chains is different from the preferred one (viz., $\pi$), and therefore a
large separation between the defect centers would increase the energy (linearly
with increasing separation).
There exists therefore a {\it space-dependent} interaction between the
polarons,
which is repulsive at short range but attractive at some
($t_{\perp}$-dependent) intermediate range. The bipolaron size, as well as
its dimensionality, depends on $t_{\perp}$ (as well as on $U$ and $V$).
There is currently limited analysis of 2D large bipolarons in the
strongly correlated limit, although some results suggest
that these can indeed be mobile
\cite{Aubry99}. Within this scenario, superconductivity occurs due to the
condensation of these large bipolarons, which is {\it not} precluded 
by the theoretical analysis of Ref.~\onlinecite{crf}.
Resolving the question of whether static distortion is sufficient,
or whether dynamical
phonons will have to be included, will require further work.

Second, in many existing models of superconducting pairing involving
correlated electrons, the interactions that bind two particles also lead to
phase separation, since the attraction producing pairing does not
saturate. Perhaps the best known example of this is 
the t-J model \cite{Dagotto,tJ} away from 1/2-filling. In contrast, within
any ``negative U'' model there does exist a saturation in this attraction
(since a single site can at most have two electrons), and the analogy
between our BCDW model and the effective 1/2-filled
``negative U -- positive V'' case suggests that
phase separation will also not occur here.  
Further, the immediately previous
discussion of the proposed binding mechanism makes clear that
with small but macroscopic (say,
1\%) concentration of commensurability defects, there is no particular
energetic advantage in creating additional polarons or bipolarons
proximate to the original bipolaron (in contrast to, say, the t-J model,
where there {\it is} such an energetic advantage).

Third, what symmetry do we expect for the superconducting order parameter in
our model?  This is clearly a challenging issue, particularly
since even with the {\it same} BCDW background 
the pairing symmetry in the highly anisotropic TMTSF
{\it might}
be different from that in the more two-dimensional BEDT-TTF and BETS.
Several recent experiments have presented evidence consistent with nodes in the
superconducting gap function in the BEDT-TTF
\cite{Slichter,Nakazawa1,Belin,Schrama}. This is reminiscent of d-wave symmetry
of the superconducting order parameter
in the high temperature copper oxide based
superconductors. On the other hand, Lee at al. have recently
presented evidence \cite{Lee1,Lee2}
suggesting that a spin triplet p-wave pairing is necessary
to explain data in (TMTSF)$_2$PF$_6$, where the
upper critical field H$_{c2}$ 
shows no saturation with the field in the plane of the organic molecules
and exceeds the Pauli paramagnetic (Clogston) limit expected
to hold for singlet superconductors \cite{Lee1}
and the temperature dependent Knight
shift measurements of $^{77}$Se show that the spin susceptibility remains 
unaltered through the superconducting T$_c$\cite{Lee2}.
Within the continuum RG theories \cite{Solyom,Emery} 
triplet superconductivity does indeed
occur proximate to the SDW. However, within 
the discrete extended Hubbard model, 
triplet superconductivity occurs within a very narrow region of the
the ``positive $U$ -- negative $V$''
sector of the $U - V$ phase diagram, bounded by the SDW phase and a 
phase segregated phase \cite{Illinois}. 
Triplet pairing thus will not only require a change in
sign of the nearest neighbor Coulomb interaction within our original
Hamiltonian of Eq.~(1), but will also occur for a very narrow critical range
of this parameter. But to resolve definitively the issue of the
symmetry of the order parameter within our model will be a non-trivial
task, as the consequences of the
interplay between e-e and e-ph interactions, as 
well as the effects of
anisotropy, must be properly understood.

\section{Acknowledgments} 

We thank Jim Gubernatis and Shiwei Zhang for discussions regarding
the CPMC method, Eduardo Fradkin and Philip Phillips for discussions
of their recent theoretical results, and Stuart Brown, Paul Chaikin,
and Andrew Schwarz for discussions of their experiments. 
We also acknowledge stimulating discussions 
with Zlatko Tesanovic.
Work at the University of Illinois
was supported in part by the grants NSF-DMR-97-12765
and NSF-GER-93-54978 and by an allocation of supercomputer
time through the NRAC program of the NCSA.

\section*{Appendix 1: The AFM-singlet transition for weak anisotropy}

Our goal here is to understand the second AFM-to-singlet 
transition that should occur in the quarter-filled band for large $t_{\perp}$
from a perspective that is different from the one presented in section III.
Specifically, we refer to the antiferromagnetic dimer lattice of 
Fig.~\ref{eff_half}(b)
with weak intrachain interdimer links, and the frozen valence bond state 
of Fig.~\ref{eff_half}(c), in which one of the interdimer links 
(W$'$ in the notation
of section III) is now stronger than the other (W in the notation of section
III), and is a singlet bond. We aim to give variational arguments at the 
simplest level that (a) point out the difference between  
$\rho$ = 1 and $\rho$ = 1/2, and (b) indicate that the frozen valence 
bond state of Fig.~~\ref{eff_half}(c)
dominates over the antiferromagnetic dimer lattice of 
Fig.~\ref{eff_half}(b) for large $t_{\perp}$ and therefore
the dimerization of the dimer lattice is unconditional. The argument
given below is not to be considered as a proof, but rather, it provides
convincing physical motivation for the numerical work discussed in section
IV. 

Note that our discussion here is limited to the relative stabilities of
two insulating states, and not the competition with any metallic state.  
We consider only the
simple Hubbard Hamiltonian with $V$ = 0 (since for $\rho$ = 1/2 the
periodicity of the CDW is the same for all $V < V_c$ and while for $\rho$ = 1
the $V$ merely reduces the effective on-site correlation) for $t_{\perp}$ = 1. 
For completeness
we begin by repeating the variational argument for the dominance
of the SDW over the BOW in $\rho$ = 1.
Consider the Heisenberg antiferromagnetic spin Hamiltonian
\begin{equation}
H = J \sum_{\langle ij \rangle} S_i . S_j
\label{Heisenberg}
\end{equation}
Consider also the singlet variational state (1,2)(3,4)....(N-1,N), with
singlet bonds between nearest neighbors in 1D and the N\'eel state
${...\uparrow \downarrow \uparrow \downarrow ...}$. The energy of an
isolated singlet bond is --(3/4)J while that of a 2-site N\'eel state is
--(1/4)J. The overall variational energy of the singlet state in 1D is
--(3/8)NJ and that of the N\'eel state --(1/4)NJ, so that the singlet dominates
over the N\'eel state in 1D. In the 2D isotropic N $\times$ N lattice, we
compare (1) the frozen valence bond state in which each chain still has the
same spin couplings as in 1D (note that at the level of our approximation
the relative phases between consecutive chains make no difference),
and (2) the 2D
N\'eel state. The variational energy of the frozen valence bond state is
--(3/8)N$^2$J, but now because of the larger number of nearest neighbors the
energy of the N\'eel state has a lower value --(1/2)N$^2$J, which therefore
dominates over the frozen valence bond state. Thus for $\rho$ = 1 in 1D
the SP state dominates, while in 2D the SDW wins over the SP state.
While this argument may appear simplistic, it nevertheless predicts the
dominance of the antiferromagnet over the singlet in $\rho$ = 1.

Consider now the isotropic 2D {\it dimerized} $\rho$ = 1/2 lattice with
moderately strong dimerization (Fig.~1(b)). 
The effective 1/2-filled band is clearly
a SDW, with the dimerization pattern being necessarily ``in-phase'' between
consecutive chains, as shown in Fig.~\ref{eff_half}(b) 
(to prevent confusion
in what follows we have not shown the bonds in Fig.~\ref{eff_half}(b), but
a strong bond between the two sites within the parentheses and weaker
interdimer bonds have been assumed)
The individual site populations are equal in this state
and each is exactly 1/2. Our contention
is that this state has a {\it higher} variational energy than that reached
by further dimerization of the dimer lattice, which gives the $\rho$ = 1/2
frozen valence bond state shown in Fig.~\ref{eff_half}(c), where there occur
interdimer singlet bonds and site occupancies ...1100... (the singlet bonds
in Fig.~\ref{eff_half}(c) are between the ``occupied'' sites). The reason for
this is that unlike in $\rho$ = 1, the exchange
integrals that describe the effective Heisenberg models in the SDW and the
singlet are now {\it different}, in spite of the fact that both Heisenberg
systems are derived from the
same Hubbard Hamiltonian. In
Fig.~\ref{eff_half}(c), we are considering
isolated singlet bonds, with site occupancies of 1,
and $J$ is clearly 2$t^2/U$, exactly as for $\rho$ = 1. In
Fig.~\ref{eff_half}(b), on the
other hand, the exchange integral has to correspond to a true $\rho$ = 1/2
system, since each site occupancy is now 1/2. The exchange integral J$'$ for
arbitrary $\rho$ in 1D is
2(${t^2/U) \rho [1 - sin (2\pi \rho/)2 \pi \rho]}$ \cite{Klein}, so that for
$\rho$ = 1/2 we have J$'$ = (1/2)J along each chain (the x-direction).
This expression is strictly true
only in the 1D undistorted chain, and for the distorted 1D chain or
in 2D one needs to calculate J$'$ from comparing
singlet-triplet
gaps within the structure corresponding to Fig.~\ref{eff_half}(b) and
within the 1/2-filled band. We have calculated
these gaps for finite lattices separately for the longitudinal and
transverse directions and
have found that while J$'$ = (1/2)J is quite accurate for the longitudinal
direction, the J$'$ in the transverse direction is {\it even smaller} (the
difference between the longitudinal and transverse directions originates
from dimerization along the longitudinal direction only), with the restriction
that only interdimer hops lead to spin exchange.
Even if we consider the largest possible value for J$'$ = J/2, the
variational energy of the N\'eel
state
in Fig.~\ref{eff_half}(b)
is then --(1/2)(N$^2$/2)J$'$ = --(1/8)N$^2$J, while that of the
frozen valence bond state in Fig.~\ref{eff_half}(c)
(with N$^2$/4 singlet bonds) is --(3/4)N$^2$/4J
= --(3/16)N$^2$J. Thus the frozen valence bond state
dominates over the dimer
SDW, implying that the dimerization of the dimer lattice is unconditional,
and the difference from the simpler $\rho$ = 1 case arises from the smaller
(by factor of 2) exchange integral in the uniform dimer lattice of
Fig.~\ref{eff_half}(b).
The above approach is obviously simplistic, but no more so than the physical
argument for the dominance of the SDW in $\rho$ = 1.

\section*{Appendix 2: Numerical Methodology}

\subsection{Contstrained Path Monte Carlo (CPMC)}

The CPMC ground-state quantum Monte Carlo method \cite{CPMC} uses a
constraining trial wavefunction to eliminate exponential
loss of signal due to the Fermion sign problem.
Although the method has been thoroughly bench-marked
against known results for the Hubbard model,
the method is non-variational, and it is important
to check its accuracy in every new system
against exact results and to use a variety of
different trial wavefunctions.  
The current application is different from previous ones in
including the $V$ interaction, as well as in the choice of the bandfilling
(previously tested cases were for bandfillings close to 1/2).
Furthermore, previous work has shown that most accurate results
are obtained when the trial noninteracting wavefunctions have ``closed-shell''
nondegenerate configurations. In the next subsection it is shown that the
proper boundary conditions for simulating coupled 1/4-filled band chains
involves having 4n electrons (where n is an integer) per chain. This implies
degeneracy of the trial wavefunctions at $t_{\perp} = 0$ and again near
$t_{\perp}$ = 1. It is thus necessary to check the accuracy of the method
for our purpose, and this was done by comparing CPMC results with the exact
results for the 8 $\times$ 2 lattice. 

\begin{figure}[htb]
\centerline{\epsfig{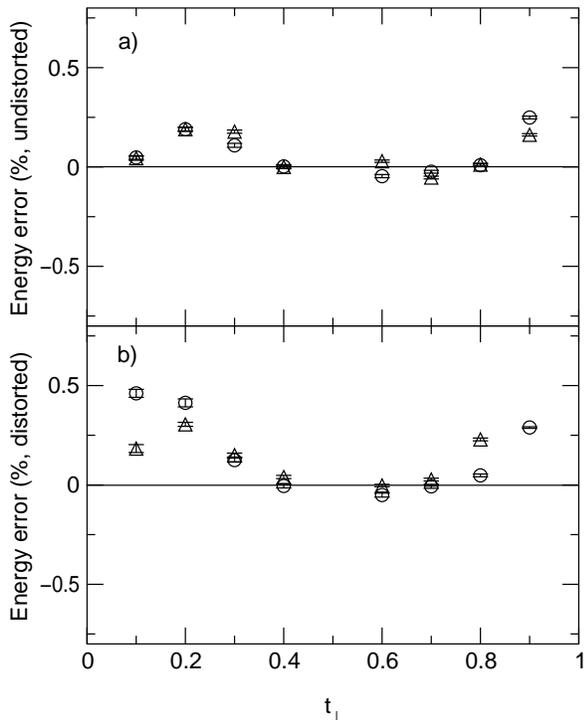}}
\caption{Percent errors in the CPMC energies for (a) undistorted and
(b) 2k$_F$ bond-distorted (the hopping integrals in the distorted
lattice correspond to those in Fig.~\ref{cartoon}(a)) 8$\times$2 lattices
with $U=6$ and $V=1$.
Triangles
are for the free-electron trial function; circles for the UHF
trial function.}
\label{cpmc_8x2_perc}
\end{figure}

Fig.~\ref{cpmc_8x2_perc} summarizes the results of the
bench-mark energy calculations for an 8$\times$2 lattice, periodic in the
x-direction,  with $U=6$ and
$V=1$. Both undistorted and
and the 2k$_F$ ($r_{2k_F}\neq 0, r_{4k_F}=0$ in Eq.~(7))
bond distorted systems were compared, where for the 
the uniform lattice all hopping
integrals were taken to be 1.0, while for the distorted system they were
1.14, 1.0, 0.86 and 1.0 (as in Fig.~2(a)).
For this amplitude of the 2k$_F$ distortion,
the absolute value of $\Delta E$ is only 0.3\%
of the total energy (at $t_\perp=0.4$). Such a
small energy difference is not easy to measure
within quantum Monte Carlo.
We note that energy differences of
this order of magnitude have also been calculated
using CPMC to study hole binding in the the 3-band Hubbard model
\cite{Guerrero}.
The CPMC values are scaled for $\Delta\tau\rightarrow 0$
from $\Delta\tau =0.05$ and $\Delta\tau =0.1$ to remove the
Trotter discretization error. The trial wavefunctions used were either
the free-electron wavefunction, or an Unrestricted Hartree-Fock
(UHF) wavefunction with $U=2$
and $V=0.5$. Hartree-Fock wavefunctions with larger $U$ and $V$
gave less accurate 
results, probably due to the tendency of UHF
to exaggerate AFM correlations.
In Fig.~\ref{cpmc_8x2_perc} the UHF trial functions produced
larger errors than the free-electron trial functions
for the distorted system at small $t_\perp$
because the SDW correlations there are exaggerated by the UHF
approximation.
The CPMC systematic errors are largest at
small $t_{\perp}$ ($<0.2$) and large $t_\perp$ ($>0.8$)
possibly due to the degeneracies in the one-electron occupancies
at $t_\perp=0$ and $t_\perp=t$.
However, at large $t_\perp$, the UHF trial wavefunction
produced slightly more accurate results for the
8$\times$2 distorted lattice possibly because the
numerically-derived UHF wavefunction breaks some
of the symmetry of the non-interacting wavefunction.
In the intermediate $t_\perp\sim 0.4$ regime,
the CPMC energies are indistinguishable from the exact energies
within the statistical error. The accuracy of the CPMC method
in this region is very reassuring, since for the {\it non}interacting
case, at  $t_{\perp}=0.4$ the distortion has already vanished.

In addition to comparing energies, we have also compared charge densities and
spin-spin correlation functions. 
Table I compares the charge densities
and spin-spin correlations computed by CPMC for the 8$\times$2
distorted lattice at $t_\perp=0.4$. The
agreement with the exact result is not as good as for the
energy (typically 1-5\% for
the charges and 5-10\% for the spin-spin correlations), but
is more than adequate to identify the presence and
periodicity of the broken symmetry states.
Thus in general, we find the CPMC results are close to the exact
results 
\begin{center}
\setlength{\tabcolsep}{0.1in}
\begin{tabular}{|c|c|c|c|c|c|}\hline
\multicolumn{3}{|c|}{$\langle\rho_j\rangle$}&
\multicolumn{3}{|c|}{$\langle s^z_i s^z_j\rangle$}\\ \hline
j       & exact         & CPMC    & i,j & exact & CPMC \\ \hline\hline
1       &0.4799         &0.4756(6)& 1,9 & -0.06095 & -0.0585(7) \\
2       &0.5201         &0.5250(6)& 1,10& -0.03215 & -0.0312(7) \\
3       &0.5201         &0.5240(6)& 1,11& 0.01408 & 0.0161(7) \\
4       &0.4799         &0.4772(6)& 1,12&-0.02698 & -0.0231(6) \\
        &               &         & 1,13&-0.07299 & -0.0687(6) \\
        &               &         & 1,14&-0.03085 & -0.0268(5) \\
        &               &         & 1,15&0.01408 & 0.0158(6) \\
        &               &         & 1,16&-0.02552 & -0.0239(7) \\\hline
\end{tabular}
\end{center}
\label{cpmc_8x2_corr}
Table I. \small Comparison of CPMC and exact charge density and spin-spin
correlations for an 8$\times$2 system with $U=6$, $V=1$, t$_\perp$=0.4, with
the same distortion of hopping integrals as in Figure~(\ref{cpmc_8x2_perc}). 
Sites on the first chain are numbered 1 -- 8, those on the second chain
9 -- 16. \normalsize\medskip\newline
for both energies and correlation functions, except for
very small or large $t_\perp$.

\subsection{Boundary Conditions}

As noted above, we determine the proper combinations
of lattices and boundary conditions for the numerical simulations
by the requirement that
nonzero $t_{\perp}$ destabilizes the BCDW for {\it noninteracting}
electrons with those boundary conditions on that particular finite lattice:
{\it i.e.}, we require the finite lattices to reflect correctly the known
behavior of the noninteracting case in the thermodynamic limit.

\begin{figure}[htb]
\centerline{\epsfig{width=3.0in,file=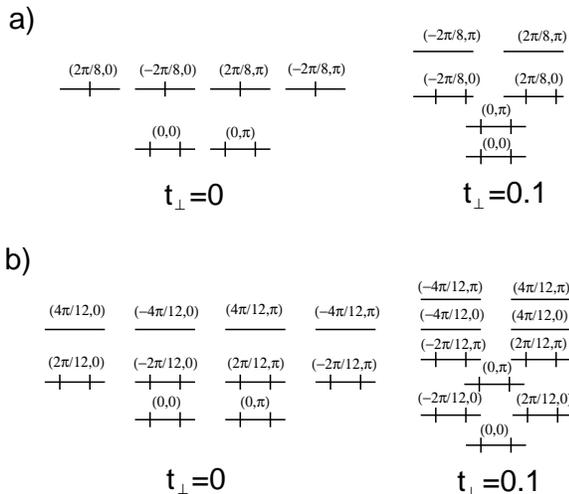}}
\caption{Occupancies of the one-electron levels for the undistorted
(a) 8 $\times$ 2 lattice, with $t_{\perp}$ = 0 (left) and $t_{\perp}$ = 0.1
(right) and (b) 12 $\times$ 2 lattice, also with
$t_{\perp}$ = 0 (left) and $t_{\perp}$ = 0.1 (right).}
\label{noninter_bands}
\end{figure}

Consider an  N $\times$ M lattice, with
N sites along the chain and M chains. To avoid odd/even
effects, consider an even number of electrons per chain.
This number can then be either 4n or 4n + 2, where n is an integer.
To obtain a 1/4-filled band, one can then have N = 4n $\times$ 2 = 8n or
N = (4n + 2) $\times$ 2= 8n + 4. The proper N for our purpose is
N = 8n ({\it i.e.}, 4n electrons per chain).
This follows from the one-electron energy levels of coupled chains
with 4n and 4n + 2 electrons per chain. In Fig.~\ref{noninter_bands}
below we have shown the one-electron
energy levels for the undistorted 8 $\times$ 2 (top panel, labeled a) and
12 $\times$ 2 (bottom panel, labeled b)
lattices (both periodic in the x-direction),
corresponding to $t_{\perp}$ = 0 on the left and 0.1 on the right in both
cases. In the 8 $\times$ 2 lattice, the degeneracy at $t_{\perp}$ = 0
will lead to spontaneous distortion.
For nonzero $t_{\perp}$ and a $\pi$-phase shift
between chains (which gives lower energy than
phase shifts of 0 or ${\pi\over 2}$), the pairs of one-electron levels that are
coupled 
by phonons with wave-vector (2k$_F$, $\pi$) are
(-- ${2\pi\over 8}$, 0) and (+ ${2\pi\over 8}$, $\pi$); and
(+ ${2\pi\over 8}$, 0)
and (-- ${2\pi\over 8}$, $\pi$). The finite gap
that occurs for $t_{\perp} \neq$ 0
between each pair of one-electron 
\begin{figure}[htb]
\centerline{\epsfig{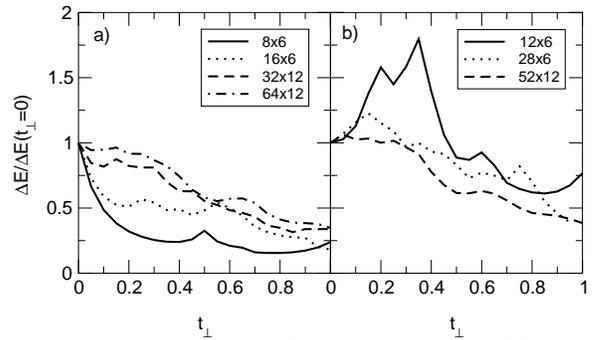}}
\caption{One-electron $\Delta$E/$\Delta$E$_0$
versus $t_{\perp}$ for (a) N = 8n and (b) N = 8n+4. In each case results
for several N $\times$ M lattices are shown.}
\label{delta_e_4n_4np2}
\end{figure}
levels coupled by the (2k$_F$, $\pi$)
phonon indicates absence of nesting
and the destabilization of the distortion. This
energy gap increases
with $t_{\perp}$, leading to a decrease in $\Delta$E
with $t_{\perp}$ for N = 8n (see Fig.~\ref{delta_e_4n_4np2} for details),
as occurs in the thermodynamic limit. In contrast, consider the
12 $\times$ 2 
lattice, in which the one-electron ground state is
non-degenerate.
There is now
a nonzero energy gap between the levels coupled by the 2k$_F$
electron-phonon interaction already at $t_{\perp}$ = 0
($k_x$ = -- ${2\pi\over 12}$ and
$k_x$ = + ${4\pi\over 12}$;
$k_x$ = + ${2\pi\over 12}$ and $k_x$ = -- ${4\pi\over 12}$). With
nonzero $t_{\perp}$, and once again a $\pi$-phase shift between the chains,
the energy gap between the levels (-- ${2\pi\over 12}$, $\pi$) and
(+ ${4\pi\over 12}$, 0),
and similarly that between the levels (+ ${2\pi\over 12}$, $\pi$) and
(-- ${4\pi\over 12}$, 0), {\it decreases}, indicating that the tendency to
distort here {\it increases with inter-chain coupling},
at least for small to moderate $t_{\perp}$.

For large N, the difference between N = 8n and
N = 8n+4 vanishes, as is shown in Fig.~\ref{delta_e_4n_4np2}, where
Figs.~\ref{delta_e_4n_4np2} (a) and (b) show the behavior of 
$\Delta E(t_{\perp})$ for N = 8n
and 8n+4, respectively.
The qualitative behavior
(destabilization of the distortion) is the same for all N = 8n,
and monotonically decreasing $\Delta$E is also seen for
N = 8n+4 for large N, but finite size effects (increasing
$\Delta$E at small 
\begin{figure}[htb]
\centerline{\epsfig{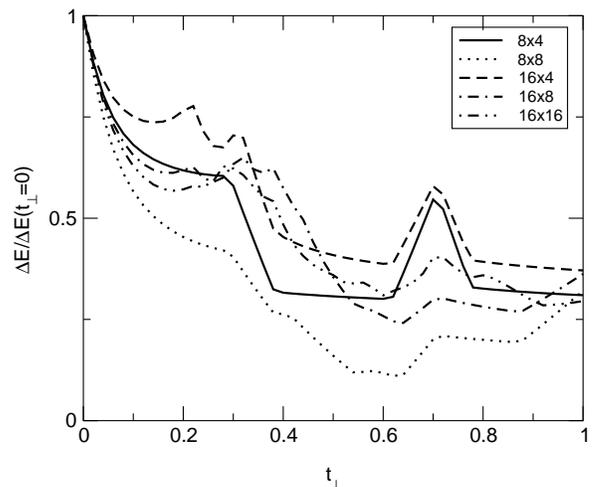}}
\caption{One-electron $\Delta$E/$\Delta$E$_0$ for the 8 $\times$ 4,
8 $\times$ 8,
16 $\times$ 4, 16 $\times$ 8 and 16 $\times$ 16 lattices
for the 2k$_F$ bond distortions as in
Fig.~\ref{cartoon}(a).}
\label{m_even}
\end{figure}
to intermediate $t_{\perp}$) are strong even
for N = 28, a chain length already too large for accurate 2D many-body
calculations. The correct qualitative behavior of all N = 8n is
the basis of our choice of these N.

In contrast to the choice of N, there is no immediate
restriction on the choice of
M, the number of chains, except that M should be even, to avoid
even/odd effects. M = 4n and 4n + 2
both show the same qualitative behavior,
as seen from the plots of $\Delta$E versus $t_{\perp}$ in
Fig.~\ref{m_even},
for several M = 4n lattices (M = 4n + 2 are included in 
Fig.~\ref{delta_e_4n_4np2}).
Thus both M = 4n and 4n + 2 are appropriate. Our choice of M = 4n + 2
is based on two reasons. First, exact diagonalization calculations on the
8$\times$2 lattice allows comparisons to results obtained within CPMC,
and the exact diagonalizations
cannot be done for the next larger appropriate lattice, viz. 8 $\times$ 4.
Second, the M = 4n lattices are characterized by one-electron Fermi
level degeneracies for
$t_{\perp}$ $\neq$ 0 (even though the degenerate levels are not coupled
by (2k$_F$, $\pi$) phonons), and the absence of a single
well-defined one-electron wave-function would make the CPMC
calculations considerably more difficult than for M = 4n + 2 lattices,
which have non-degenerate one-electron levels for nonzero $t_{\perp}$.

\subsection{UHF calculations of bond distortion}

As discussed in the subsection on methods in this Appendix, UHF trial 
wavefunctions for the CPMC calculations were constructed for regions where
one-electron wavefunctions were degenerate. Since UHF calculations give
reasonably correct results in the small $U, V$ range it is also of interest to
determine the tendency of the 2D lattice to distort within the UHF 
approximation. One advantage of this procedure is that much larger lattices
than those discussed in section IV can be tested. We report these results here.
We have chosen
relatively small $U$ and $V$ for
two reasons: the UHF procedure does not 
converge well for
larger interactions, and the smaller 
\begin{figure}[htb]
\centerline{\epsfig{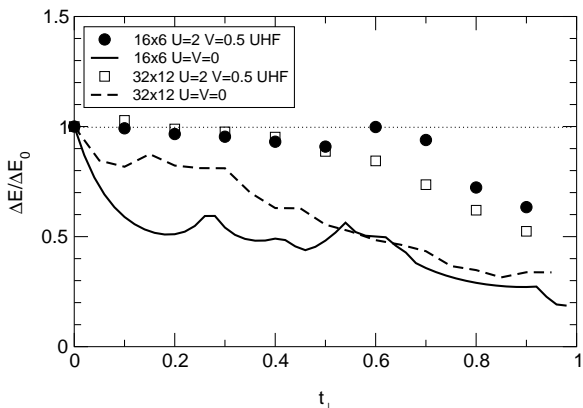}}
\caption{$\Delta E/\Delta E_0$ versus $t_\perp$ for a 2k$_F$
bond distortion ($r_{4k_F}=0$)
for noninteracting and the interacting
lattices within the UHF approximation. Intra-chain hopping
integrals for the distorted lattices
are as indicated in Fig.~\ref{cartoon}(a).}
\label{hf_delta_e}
\end{figure}
values of $U$ and $V$
gave better results when used as a CPMC trial function
(compared to a numerically exactly solved
 8$\times$2 system).
Fig.~\ref{hf_delta_e} shows the normalized energy gain
from a 2k$_F$ distortion for two different lattices, within the
UHF approximation. The UHF results show that
$\Delta E/\Delta E_0$ remains close to 1 for at least up to
$t_{\perp} \sim$ 0.4, indicating a tendency to persistent distortion up to
this $t_{\perp}$. Although
$\Delta E/\Delta E_0$ begins to decrease at still larger $t_{\perp}$,
these calculations are for a
relatively small value of $U$, and as discussed in section III, the range
of $t_{\perp}$ over which the distortion should persist increases with $U$.
Thus the qualitative effects of the e-e interaction are already visible within
the UHF approach at small $U$, while a fully persistent broken symmetry state
will occur only for larger values of the e-e interaction that are beyond
the scope of the UHF.
Given that the UHF approximation predicts a vanishing of the bond dimerization
in the 1/2-filled band for a fairly small $U_c$ (the actual magnitude of
$U_c$ depends on $\alpha$), in contrast to the correct result that there
is an enhancement
of the dimerization \cite{review1} for $0 < U < 4$, the present results,
showing a persistence of the distortion for moderate $t_{\perp}$, is
initially perplexing. 
The reason for the correct prediction
in this case
is that the UHF exaggerates the
SDW, which destroys the BOW in the 1/2-filled band, but has a co-operative
interaction with the 1/4-filled band BOW for small to moderate $t_{\perp}$.

\section*{Appendix 3: Spin character of the ground state}

As discussed in Appendix 2,
the proper boundary condition for the numerical
evaluation of the electronic energy gained upon bond or site distortion in
$\rho$ = 1/2 involves finite N $\times$ M lattices with N = 8n.
This requires the number of electrons per chain to be 4n, and it is known
that in 1D periodic {\it undistorted} rings with $\rho \neq$ 1, the ground
state has 
overall spin S = 1 instead of 0 for any nonzero $U$. 
\setlength{\tabcolsep}{0.05in}
\begin{center}
\begin{tabular}{|l|l|l|l|l|} \hline 
$t_{\perp}$& \multicolumn{2}{c|}{undistorted} &\multicolumn{2}{c|}{2k$_F$
distortion} \\ \hline
& S=0& S=1&S=0&S=1 \\ \hline
0.01 &-9.335651&-9.335637&-9.352522&-9.352228 \\
0.025&-9.337570&-9.336944&-9.354380&-9.353739\\
0.05&-9.344122&-9.341546&-9.361083&-9.358425\\ \hline
\end{tabular}
\end{center}
Table II: \small The S=0 and S=1 energies of the 8$\times$2 
undistorted and 2k$_F$ bond-distorted lattice for
$U = 6$ and $V = 1$. The lowest energy is S=0 for both
undistorted and distorted cases. \normalsize\medskip\newline
The spin of the ground state of the distorted periodic ring depends on 
its size and
the magnitude of the Hubbard $U$. For the values
of the correlation parameters and bond distortion parameter in 
Fig.~\ref{delta_e_2kf}, the 
ground state in the N = 8 distorted periodic ring has S = 1, while the
N = 16 ground state has S = 0. Thus the $\Delta E_0$ in 
Fig.~\ref{delta_e_2kf} for
nonzero e-e interaction corresponds to $\Delta E_{TT}$ ({\it i.e.}, the energy
gained by the triplet state upon bond distortion) for N = 8, and to
$\Delta E_{TS}$ (undistorted state in S = 1, distorted state in S = 0)
for N = 16. Whether or not the comparisons of the zero and nonzero $t_{\perp}$
are then meaningful is an important question.
We present here the detailed results of three different sets of calculations,
each of which indicates that our interpretation of the results of 
Fig.~\ref{delta_e_2kf}
(viz., strong tendency of the interacting 1/4-filled
lattice to distort
at arbitrary $t_{\perp}$) is correct.

First, we have calculated the exact ground states of the 8 $\times$ 2 lattice
for $t_{\perp}$ as small as 0.01. 
In Table II we have
given the S = 0 and S = 1 energies of the 8 $\times$ 2 lattice for
$U$ = 6 and $V = 1$, for three small values of $t_{\perp}$.
The coupled chain system is in the S = 0 state for {\it both}
zero and nonzero bond
distortion for the smallest nonzero $t_{\perp}$. The important point now is
that instead of choosing the single isolated chain as the
standard in Fig.~\ref{delta_e_2kf}, we could have also chosen the coupled chain system
with $t_{\perp}$ = 0.01 as the standard, provided the distortion of the
$t_{\perp}$ = 0.01 lattice is also unconditional. Even if the nesting
ideas were valid, we believe that the coupled chain system with 
$t_{\perp}$ = 0.01 is unconditionally distorted
and then the results in Table II clearly show
that $\Delta E$ increases with further increase in $t_{\perp}$, indicating
enhanced distortion relative to $t_{\perp}$ = 0.01. The error bars
in the CPMC calculations prevent us from performing similar calculations
for the 8 $\times$ 6 or the 16 $\times$ 6 lattices, but the overall 
similarities in the (a) occupancies of the one-electron levels for
nonzero $t_{\perp}$ and (b) $\Delta E$ behavior, especially in the region
$t_{\perp} \leq 0.4$,
preclude different behavior at small nonzero $t_{\perp}$.

We performed a second set of calculations for the 8 $\times$ 2 lattice for
very small values of $U$ (with $V$ = 0). Note that if the persistent
distortion implied in Fig.~\ref{delta_e_2kf} were merely due to our 
choosing the wrong
reference point $t_{\perp}$ = 0 
\begin{figure}[htb]
\centerline{\epsfig{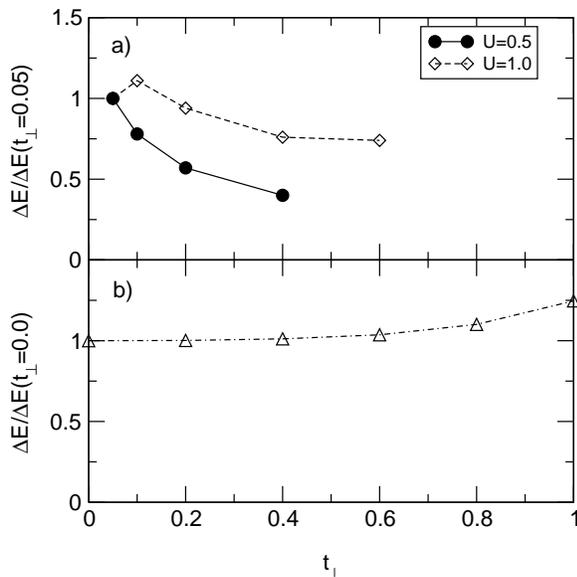}}
\caption{(a) $\Delta E$ {\it vs.} $t_{\perp}$ for the 8$\times$2
lattice at small $U$ (normalized to the value at $t_{\perp}=0.05$). Note
the decrease in the $\Delta E$.
(b)$\Delta E/\Delta E_0$ {\it vs.} $t_{\perp}$ for the 8$\times$2
lattice at $U=100$.}
\label{app-b}
\end{figure}
(since exactly at this point
$\Delta E_0$ = $\Delta E_{TT}$), an apparently enhanced distortion for
nonzero $t_{\perp}$ should
occur for {\it all} nonzero $U$ (since the single chain is S = 1 for all
nonzero $U$, while the coupled chain system has S = 0 for all nonzero
$t_{\perp}$ and $U$). 
On the other hand, if the results 
in Fig.~\ref{delta_e_2kf}
are due to the
confinement effect discussed in section III.D, then
enhanced/persistent distortion
should occur only {\it above a threshold e-e interaction}: for weak
e-e interaction the behavior should resemble that of the noninteracting
lattice (with enhanced or persistent distortion occurring for a small range
of $t_{\perp}$ near $t_{\perp}$ = 0). We show here the results of calculations
at small $U$ for {\it site}
distortion (as opposed to bond distortion), since we also report calculations
for very large $U$ below, and the bond distortion pattern (the magnitude of
$r_{4k_F}$) is $U$-dependent, but the site distortion pattern is not.
The distorted lattice here has site energies + $\epsilon$,
+ $\epsilon$, -- $\epsilon$, -- $\epsilon$ (with $\epsilon = 0.1$)
over four consecutive sites, and a $\pi$-phase shift between the two 
periodic rings.
Since the 2k$_F$ CDW has a synergetic coexistence with both the r$_{4k_F}$
= 0 BOW (Fig.~\ref{cartoon}(a)) and the r$_{4k_F} \neq$ 0 BOW 
(Fig.~\ref{cartoon}(b)) \cite{umt}
a persistent CDW
also implies persistent BOW; we have confirmed this by calculating the
expectation values of the bond orders. 
In Fig.~\ref{app-b}(a) we show the $\Delta E$ behavior as
a function of $t_{\perp}$ for both $U = 0.5$ and $U = 1$. 
Decreasing
$\Delta E$ with $t_{\perp}$ is a clear signature that the 
tendency to distortion here
{\it decreases} with increasing two dimensionality, since confinement at these
small $U$ is not sufficient to give persistent distortion. Even though
these calculations are with fixed site energies, the expectation values of the
charge densities depend on $t_{\perp}$, and our calculated CDW amplitudes
decrease with $t_{\perp}$, as expected from Fig.~\ref{app-b}(a). This
behavior is exactly opposite to that in Fig.~\ref{delta_rho}(b), indicating
again a decrease in distortion with $t_{\perp}$ at small $U$.
Finally, we emphasize
that similar calculations have also been done with fixed 2k$_F$ bond
distortion, and once again we observe decreasing $\Delta E$
and CDW amplitude with increasing $t_{\perp}$.

We performed a third set of calculations with very large $U = 100$, again with 
the same site distorted lattice but now with $\epsilon$ = 0.2, since at this very
large $U$, the energy gained upon distortion for $\epsilon$ = 0.1 is very
small.
The resultant BOW here has strong 4k$_F$ component ($r_{4k_F} \neq 0$),
and this is why the distorted lattice was chosen to be 
the 2k$_F$ CDW in this and the above calculations, such that meaningful
comparisons between these extreme cases can be made.
At this
large $U$, the energy difference between S = 0 and S = 1 states is negligible.
For example, for the 1D 8-site periodic ring $\Delta E_{SS}$ (electronic
energy gained in the S = 0 subspace, with both undistorted and distorted
states in S = 0) = 0.06222, while $\Delta E_{TT}$ (electronic
energy gained in the S = 1 subspace, with both undistorted and distorted
states in S = 1) = 0.06224. Fig.~\ref{app-b}(b) shows the  $\Delta E$ behavior
as a function of $t_{\perp}$ 
(with $\epsilon$ = 0.2 now). An enhanced CDW (and therefore BOW) is seen from
as a function of $t_{\perp}$, where the singlet and triplet
data points at $t_{\perp}$ = 0
are the same. As seen in Fig.~\ref{app-b}(b), the $\Delta E$ for nonzero
$t_{\perp}$ is weakly enhanced now even when compared to $\Delta E_{SS}$ at
$t_{\perp}$ = 0. Once again, the behavior of the CDW amplitude is in complete
agreement with the prediction from Fig.~\ref{app-b}(b), viz., a weak 
enhancement of the CDW amplitude with $t_{\perp}$.

Considering the above three different sets of results, we therefore conclude 
that the results in Fig.~\ref{delta_e_2kf}, Fig.~\ref{delta_rho}(b) and
Fig.~\ref{delta_e_dd}
are not artifacts, and
the persistent distortion is real and a true confinement effect, as would
also be expected from the ``variational''
arguments in Appendix 1.


\begin{references}
\bibitem{ladder-footnote}
We exclude from our discussions the extensive work on doped 2- and 4-leg
ladder systems. It is not clear currently whether or not these results are
relevant for the true 2D lattice.
\bibitem{disorder} A true metallic phase may not be obtained because of
disorder effects, but this is a different issue that is unrelated to the
question of the existence of density waves, and is hence ignored here.
\bibitem{confinement3} Given that the issue of confinement in interacting
electron systems has been hotly debated, it is pertinent to emphasize that
we use this expression in a considerably narrower sense than has sometimes been
used in the literature. Our interest lies strictly in the interacting 1/4-filled
band case, and it is not being implied that the effect discussed here applies
to other bandfillings.
\bibitem{nomenclature} 
 In all cases X are inorganic anions that play only indirect
roles in accepting electrons from the organic cations, and do not have any
active role in charge transport, although they do influence the magnitudes
of the interchain hopping integrals.
\bibitem{xysym} The underlying $x \leftrightarrow y$ symmetry in
the isotropic 2D limit
implies that there are two
degenerate orthogonal 2D BCDW states for $t_{\perp}$ = $t$.
\end{references}
\end{document}